\documentclass{emulateapj}
\usepackage{amsmath, natbib, ulem}
\usepackage[usenames,dvips]{color}

\slugcomment{Submitted to the Astrophysical Journal}

\begin{document}

\title{Interacting Binaries with Eccentric Orbits. III. Orbital Evolution due
to Direct Impact and Self-Accretion}

\author{J. F. Sepinsky$^{1,2}$, B. Willems $^2$, V. Kalogera$^2$, F.
A. Rasio$^2$}
\affil{$^1$ The University of Scranton, Department of Physics and
Electrical Engineerings, Scranton, PA 18510}
\affil{$^2$ Department of Physics and Astronomy, Northwestern
University,
2145 Sheridan Road, Evanston, IL 60208}
\slugcomment{jeremy.sepinsky@scranton.edu, b-willems, vicky, and 
rasio@northwestern.edu}
\shorttitle{Direct Impact and Self-Accretion in Eccentric Binaries}

\begin{abstract}

The rapid circularization and synchronization of the stellar components 
in an eccentric binary system at the onset of Roche lobe overflow is a 
fundamental assumption common to all binary stellar evolution and 
population synthesis codes, even though the validity of this assumption 
is questionable both theoretically and observationally. Here we 
calculate the evolution of the orbital elements of an eccentric binary 
through the direct three-body integration of a massive particle ejected 
through the inner Lagrangian point of the donor star at periastron.  The 
trajectory of this particle leads to three possible outcomes: direct 
accretion onto the companion star within a single orbit, self-accretion 
back onto the donor star within a single orbit, or a quasi-periodic 
orbit around the companion star, possibly leading to the formation of a 
disk. We calculate the secular evolution of the binary orbit in the 
first two cases and conclude that direct impact accretion can increase 
as well as decrease the orbital semi-major axis and eccentricity, while 
self-accretion always decreases the orbital semi-major axis and 
eccentricity. In cases where mass overflow contributes to circularizing 
the orbit, circularization can set in on timescales as short as a few 
per cent of the mass transfer timescale. In cases where mass overflow 
increases the eccentricity, the orbital evolution is governed by 
competition between mass overflow and tidal torques. In the absence of 
tidal torques, mass overflow results in direct impact can lead to 
substantially subsynchronously rotating donor stars. Contrary to 
assumptions common in the literature, direct impact accretion 
furthermore does not always provide a strong sink of orbital angular 
momentum in close mass-transferring binaries; in fact we instead find 
that a significant part can be returned to the orbit during the particle 
orbit. The formulation presented in this paper together with our 
previous work can be combined with stellar and binary evolution codes to 
generate a better picture of the evolution of eccentric, 
Roche lobe overflowing binary star systems.
\end{abstract}


\section{Introduction}

In close binaries, the combined effects of stellar and orbital evolution 
can cause a star to fill its Roche lobe and transfer mass to its 
companion. If the orbit is eccentric this mass transfer is expected to 
commence at or near the periastron of the binary orbit where the 
effective Roche lobes of the component stars are smallest 
\citep{1998NewA....3..111L, 2007ApJ...660.1624S, 2009MNRAS.tmp..394C, 
2005MNRAS.358..544R}. Until recently, studies of mass transfer in 
eccentric binaries mainly focused on smoothed particle hydrodynamics 
calculations of the mass transfer stream and much less on the effects of 
the mass transfer on secular binary evolution.

To the best of our knowledge, \citet{2007ApJ...667.1170S, 
2009ApJ...702.1387S} were the first to study in detail the secular 
evolution of the orbital semi-major axis and eccentricity of eccentric 
mass-transferring binaries. The authors adopted a perturbation method 
initially outlined by \citet{1969Ap&SS...3..330H} and found that, 
depending on the binary properties at the onset of mass transfer, the 
orbital semi-major axis and eccentricity can either increase or decrease 
at a rate linearly proportional the mass transfer rate at periastron. 
Contrary to common assumptions, tides therefore do not always rapidly 
circularize binary orbits as a star evolves towards filling its Roche 
lobe.
 
\citet{2007ApJ...667.1170S, 2009ApJ...702.1387S} calculated the orbital 
evolution due to mass transfer in eccentric binaries assuming mass 
transfer leads to the formation of an accretion disk. Tidal interactions 
between the disk and the donor star then allow at least some fraction of 
the angular momentum carried by the transferred matter to be returned to 
the orbit. However, if the size of the accretor is a considerable 
fraction of the size of the orbit, matter lost by the donor star through 
the inner Lagrangian point may impact the accretor's surface directly 
rather than form an accretion disk.

In this paper, we complement the work of \citet[][ hereafter 
Paper~I]{2007ApJ...667.1170S} and \citet{2009ApJ...702.1387S} by 
investigating the orbital evolution due to mass overflow in eccentric 
binaries in which no disk forms around the accretor. For this purpose, 
we calculate the trajectories of mass elements escaping through the 
inner Lagrangian point of the donor star in the ballistic limit, and 
distinguish between three possible cases: (1) the ejected matter falls 
back onto the donor star, (2) the ejected matter impacts the surface of 
the accretor, and (3) the ejected matter orbits the accretor for at 
least one orbital revolution of the binary.  In the first two cases, the 
evolution of the orbital elements is obtained directly from the 
ballistic trajectory calculations. In the third case, the evolution of 
the orbital elements depends on the interactions between mass elements 
ejected during successive periastron passages of the donor star. Since 
such interactions are expected to violate our assumption of particles 
moving in the ballistic limit, we do not consider them here.

The present paper is organized as follows.  In \S\,2, we outline the 
basic assumptions and methodology used to study ballistic particle 
trajectories and the effects of mass overflow on the orbital evolution 
of eccentric binaries. In \S\,3, we examine the fate of mass lost by the 
donor star as a function of the initial binary properties and identify 
ranges of initial parameters leading to fallback, direct impact 
accretion, and possible disk formation. Evolutionary sequences and 
timescales of orbital evolution due to mass overflow in eccentric 
binaries are presented in \S\,4. The final section, \S\,5, is devoted to 
concluding remarks.

\section{Mass overflow in eccentric binaries}
\label{sec-MTinecc}

\subsection{Basic assumptions}

We consider a close binary system of stars with masses $M_1$ and $M_2$, 
and spherical radii ${\cal R}_1$ and ${\cal R}_2$. We define the binary 
mass ratio as $q=M_1/M_2$, and assume the initial orbit to be Keplerian 
with semi-major axis $a$, eccentricity $e$, and orbital period $P_{\rm 
orb}$. The periastron distance of the two stars is then $r_P~=~a(1-e)$.

The binary component stars are assumed to rotate uniformly with 
rotational angular velocities $\vec{\Omega}_1$ and $\vec{\Omega}_2$ 
around axes perpendicular to the orbital plane and in the same sense as 
the orbital motion.  For convenience, we express the rotational angular 
velocities in units of the orbital angular velocity $\vec{\Omega}_P$ at 
periastron:
\begin{equation}
 \vec{\Omega}_1 = f_1\,\vec{\Omega}_P, \,\,\,\,\,\,\,\,\,\, 
 \vec{\Omega}_2 = f_2\, \vec{\Omega}_P
\end{equation} 
with
\begin{equation}
|\vec{\Omega}_P| = \frac{(1+e)^{1/2}}{(1-e)^{3/2}}\, 
\frac{2\,\pi}{P_{\rm orb}}.
\end{equation}
Hence, $f_1$ and $f_2$ are a dimensionless parameterization of the 
rotation rates of stars~1 and 2, respectively, in terms of 
$|\vec{\Omega}_P|$.

For stars in a circular synchronous binary orbit, it is commonly assumed 
that the surface of the star will conform to the equipotential surface 
which encloses the same volume as the spherical radius of the star.  The 
maximum size of the star is then determined by the size of the Roche 
lobe, which is defined by the equipotential surface passing through the 
$L_1$ point.  For eccentric binaries, we follow 
\citet{2007ApJ...660.1624S, 2007ApJ...667.1170S} and construct effective 
equipotential surfaces using a reference frame corotating with the star 
such that the envelope is instantaneously quasi-static \citep[see 
also][]{Plavec58, 1963ApJ...138.1112L, 1976ApJ...209..574A}.  For an 
analysis of the validity of this assumption, see 
\citet{2007ApJ...660.1624S}.  In this reference frame we can locate the 
effective $L_1$ point at the minimum of the potential along the line 
connecting the mass centers of the two stars \footnote{ We note that 
viscosity in the stellar atmosphere will cause a tidal lag or lead in 
the position of $L_1$ with respect to line connecting the centers of the 
two stars, to a degree dependent on the asynchronism of the star.  We 
intend to include tidal lags and leads in future investigations.
}.  We can then identify the effective Roche lobe as the equipotential
surface passing through this effective $L_1$ point.

At some point in time, one of the stars is assumed to fill its effective 
Roche lobe due to the expansion of the star and/or the contraction of 
the orbit.  For brevity, we refer to the effective Roche lobe-filling 
star as star~1 and to its companion as star~2.  Since the stars are in 
an eccentric orbit, the size of star 1's effective Roche lobe is a 
function of time and is smallest when the stars are at periastron 
\citep[][ hereafter SWK]{2007ApJ...660.1624S}. Roche lobe overflow is 
therefore expected to take place first at periastron and then re-occur 
at each subsequent periastron passage.

To model the episodic mass loss at successive periastron passages, we
assume a small amount of mass $M_3$ ($M_3 \ll M_1, M_2$) to be ejected
from star~1 through its inner Lagrangian point $L_1$ at each periastron
passage.  As in Paper~I, we assume the mass overflow rate $\dot{M}_1$
from the star to be a Dirac delta function of amplitude $\dot{M}_0<0$
centered on the periastron of the binary orbit:
\begin{equation}
\dot{M}_1 = \dot{M}_0\, \delta(\nu)
\label{eq-m0}
\end{equation}
where $\nu$ is the true anomaly with $\nu=0$ at the periastron, and a 
normalization factor of $2\,\pi$ has been absorbed in the constant value for
$\dot{M}_0$. The mass of the ejected particle is then related to the 
mass overflow rate $\dot{M}_0$ by
\begin{equation}
M_3 = \int^{P_{\rm orb}/2}_{-P_{\rm orb}/2} (-\dot{M}_1)\, dt = 
  -\frac{\left(1-e\right)^{3/2}}{\left(1+e \right)^{1/2}} \, 
\frac{P_{\rm orb}}{2\pi} \, \dot{M}_0,
\label{M3}
\end{equation}
where we made use of the relation
\begin{equation}
dt = \frac{\left(1-e^2\right)^{3/2}}{\left(1+e\cos{\nu}\right)^2} \, 
\frac{P_{\rm orb}}{2\pi} \, d\nu. 
\end{equation}

\begin{figure*}
\plotone{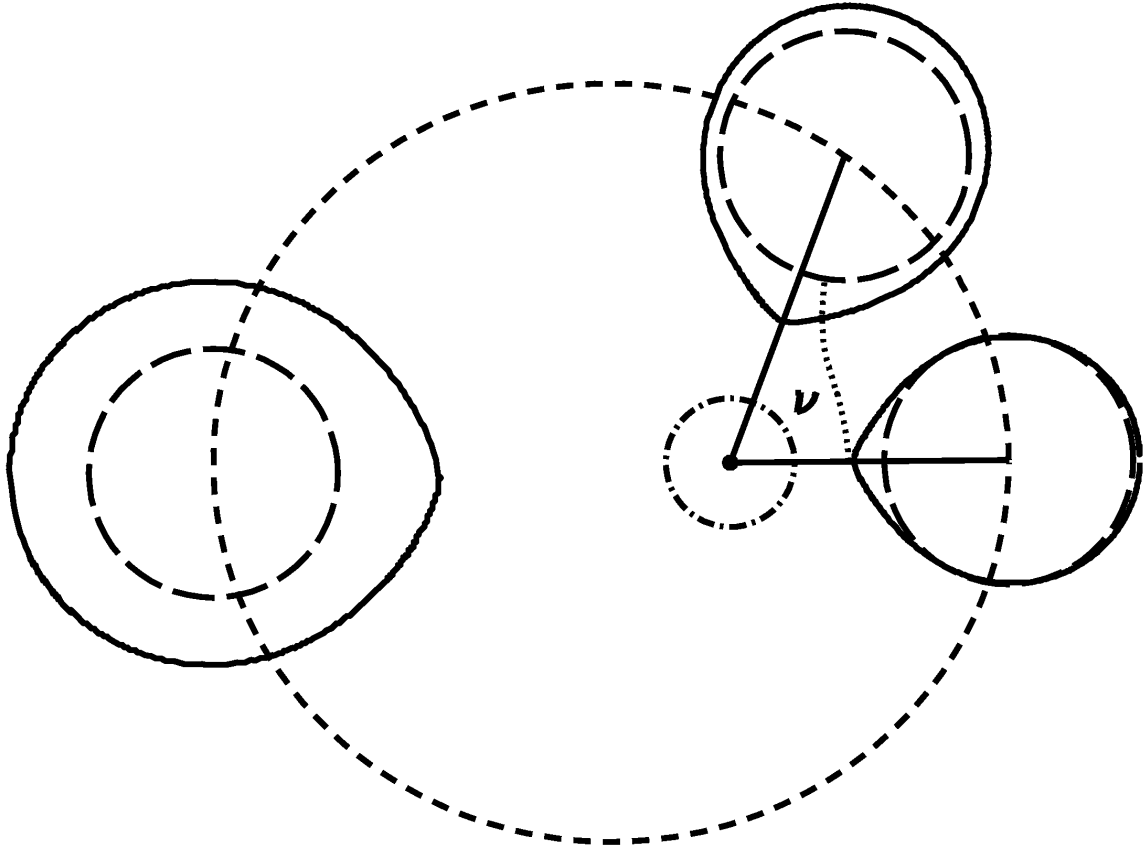}
\caption{
A schematic representation of the orbit of the donor star (short-dashed 
line) in a reference frame fixed to star~2.  In this diagram, we show 
the constant radius of star~2 (dashed-dotted lines), the constant radius 
of the donor (long-dashed line), the time-dependent effective Roche lobe 
(solid line), and the path of the ejected particle (dotted line).  The 
donor star is shown at three different positions in the orbit: 
periastron, particle impact, and apastron. The true anomaly, $\nu$, is 
the angle of the line connecting the mass centers of the stars measured 
with respect to periastron, and is indicated above.  For this orbit 
($q_i=2.0$, $f_{1,i}=0.8$, $e_i=0.3$), the ejected particle impacts the 
donor star at $\nu\simeq 1.21$ radians.
}
\label{fig-orbit}
\end{figure*}

In figure~\ref{fig-orbit}, we show the orbit for a system with initial 
eccentricity $e_i=0.3$, initial donor rotation rate $f_{1,i}=0.8$, and 
initial mass ratio $q_i=2.0$ in a reference frame where star~2 is 
stationary. The ejected particle undergoes self-accretion, and its 
trajectory is shown by the dotted line.  The orbit of the donor star is 
eccentric and shown by the short-dashed line.  The position of star~2 is 
indicated by the point, and the dashed-dotted line shows its spherical 
radius.  The donor star is shown at three positions (from right to left, 
respectively): periastron, particle impact, and apastron, and moves in 
the counter-clockwise direction.  The spherical radius of the donor 
star, ${\cal R}_1$, is shown by the long-dashed line and is a constant 
radius for each of the positions shown.  The effective Roche lobe of 
star~1 is also drawn at each position.  It is clear that the effective 
Roche lobe radius of star~1 grows relative to its initial size as it 
moves away from periastron.  We also indicate the true anomaly, $\nu$,  
which measures the angle of the line connecting the mass centers of the 
stars in a Keplerian orbit measured counter-clockwise from periastron.  

Furthermore, we consider the particle to impact the surface of a star 
when it falls within the spherical radius of that star.  This surface 
does not lie at the Roche lobe, as can be seen in 
Figure~\ref{fig-orbit}.  Due to the computational complexity and 
time-variable shape of the effective Roche lobe, calculations of this 
kind were not considered here.  If this calculation were included, it is 
clear that the particle would strike the surface of the effective Roche 
lobe earlier in the orbit.  In fact, for many of the cases studied, the 
ejected particle never moves further from the donor star than the 
position of the $L_1$ point at periastron.  Thus, for these cases, if we 
assumed impact occurred at the surface of the effective Roche lobe, no 
mass loss would occur at the $L_1$ point, and thus we would not expect 
any orbital evolution due to this effect.  As such, it may be possible 
for non-synchronous, eccentric systems to effectively halt mass loss 
through the $L_1$ point.  We will investigate this effect in a future 
study.

\subsection{Ballistic particle trajectories}
\label{sec-trajectories}

Our aim in this paper is to investigate the evolution of the orbital 
elements due to mass overflow in eccentric binaries. For this purpose, 
we treat the binary components as rigid spheres of uniform density. The 
matter ejected by the donor star at each periastron passage is treated 
as a point mass. To describe the motion of the stars and the ejected 
mass elements, we introduce an inertial frame of reference $OXYZ$ with 
origin $O$ at the initial center of mass of the binary prior to any mass 
overflow, and with the $XY$-plane coinciding with the orbital plane. The 
orientation of the axes is chosen such that the positive direction of 
the $X$-axis coincides with the line from the binary mass center to the 
position of the donor star at the onset of mass overflow, and the 
positive $Z$-axis is parallel to the direction of the orbital angular 
momentum vector at the onset of mass overflow.

To determine whether a star undergoes mass overflow, its 
volume-equivalent Roche lobe radius is calculated accounting for the 
effects of eccentricity and non-synchronous rotation, as described in 
detail in SWK. Since we here do not consider the detailed structure 
of the donor star, we assume the matter ejected at periastron to have an 
initial velocity $\vec{V}_3$ equal to the vector sum of the star's 
rotational velocity at $L_1$ and its orbital velocity at periastron:
\begin{equation}
\vec{V}_3 = \vec{\Omega}_1 \times \vec{r}_{L_1}^{} + \vec{V}_1.
\label{eq-vpi}
\end{equation}
Here, $\vec{r}_{L_1}$ is the vector from the center of mass of star~1 to
the $L_1$ point, and $\vec{V}_1$ the orbital velocity of star~1 at
periastron right before the ejection of the matter.  We do not
consider any contribution to $\vec{V}_3$ from the thermal speed
of the mass elements in the donor's atmosphere
\citep{1975ApJ...198..383L}. A generalization of our calculations
accounting for the thermal speed is straightforward, but adds  
extra dimensions to the parameter space explored in this paper. We 
therefore defer this extension to a future investigation. 

After the ejection of a mass element from the $L_1$ point, the system 
evolves as a three-body system according to the equations of motion
\begin{equation}
\frac{d^2 \vec{R}_k}{dt^2} = - 
\sum_{\scriptsize \begin{array}{c}j=1\\j 
\neq k\end{array}}^3 \frac{G\,M_j}{|\vec{r}_{kj}|^2}\, 
\frac{\vec{r}_{kj}}{|\vec{r}_{kj}|} \hspace{0.7cm} (k=1,2,3),
\label{3body}
\end{equation} 
where $G$ is the Newtonian gravitational constant, $\vec{R}_1$ and 
$\vec{R}_2$ are the position vectors of star~1 and star~2 with respect 
to the inertial frame of reference, $\vec{R}_3$ is the position vector 
of the ejected mass element with respect to the inertial frame, and 
$\vec{r}_{kj}=\vec{R}_k-\vec{R}_j$.

For numerical convenience, we express equations (\ref{3body}) and the 
initial condition given by equation (\ref{eq-vpi}) in terms dimensionless 
quantities $\vec{R}_j^*$, $M_j^*$, $t^*$, $\vec{V}_j^*$, and 
$\vec{\Omega}_P^*$ defined as 
\begin{eqnarray}
 \vec{R}_j^* &=& \vec{R}_j/r_{P,i}, \,\,\,\,\, j=1,2,3 \\
 M_j^* &=& M_j/M_T, \,\,\,\,\, j=1,2,3 \\
 \vec{V}_j^* &=& \vec{V}_j\sqrt{\frac{r^{}_{P,i}}{GM_T}}, \,\,\,\,\, 
 j=1,2,3\\
 t^* &=& t \sqrt{\frac{GM_T}{r_{P,i}^3}}, \\
 \vec{\Omega}_P^* &=& \vec{\Omega}_P \sqrt{\frac{r_{P,i}^3}{GM_T}},
\end{eqnarray}
where $r_{P,i}^{}$ is the initial periastron distance between the binary 
components at the onset of mass overflow, $M_T$ is the total system 
mass, and the subscripts $j=1,2,3$ refer to star~1, star~2, and the 
ejected particle, respectively. In equation (\ref{eq-vpi}), we also 
express the position of the $L_1$ point in units of the initial periastron 
distance as
\begin{equation}
 \vec{r}_{L_1}^{\,*} = \frac{\vec{r}_{L_1}^{}}{r_{P,i}^{}}.
\end{equation}
The quantity $\vec{r}_{L_1}^{\,*}$ depends only on the binary mass 
ratio, the orbital eccentricity, the true anomaly, and the rotation rate 
of star~1 in units of the orbital angular velocity at periastron (See 
Equation [A15] of Paper 1).

In terms of these dimensionless quantities, the initial positions and 
velocities of the stars and ejected particle are 
\begin{eqnarray}
 \vec{R}_{1,i}^* &\equiv& (X_{1,i}^*, Y_{1,i}^*, Z_{1,i}^*) = 
 (\frac{1}{1+q_i},0,0),  \label{do1} \\
 \vec{R}_{2,i}^* &\equiv& (X_{2,i}^*, Y_{2,i}^*, Z_{2,i}^*) = 
 (-\frac{q_i}{1+q_i},0,0),  \\
 \vec{R}_{3,i}^* &=& \vec{r}_{L_1,i}^{\,*} \\
 \vec{V}_{1,i}^* &=& \vec{R}_{1,i}^*\times\vec{\Omega}_{P,i}^*, \\
 \vec{V}_{2,i}^* &=& \vec{R}_{2,i}^*\times\vec{\Omega}_{P,i}^*,\\
 \vec{V}_{3,i}^* &=& f_{1,i}\vec{\Omega}_{P,i}^* \times 
 \vec{r}_{L_1,i}^{\,*} + \vec{R}_{1,i}^*\times\vec{\Omega}_{P,i}^*,
\end{eqnarray}
where the subscript ``$i$'' indicates that the quantity is to be
considered at the time the particle is ejected, recalling that the 
particle is ejected at periastron.  Furthermore, equation~\ref{3body} 
becomes
\begin{equation}
\frac{d^2 \vec{R}_k^*}{d(t^*)^2} = -
\sum_{\scriptsize \begin{array}{c}j=1\\j\neq
k\end{array}}^3 \frac{M_j^*}{|\vec{r}_{kj}^{\,*}|^2}\,
\frac{\vec{r}_{kj}^{\,*}}{|\vec{r}_{kj}^{\,*}|} \hspace{0.7cm} (k=1,2,3).
\label{do}
\end{equation}
In this form, both the initial conditions and the equations of motion are 
independent of the initial size of the orbit.  

We use an 8th order Runge Kutta ordinary differential equation 
solver \citep{GNUGSL} to integrate equations~(\ref{do}) forward in 
time, starting from the first periastron passage at which matter is 
ejected by the donor. The three-body integration ends when (1) the 
ejected matter impacts star~2 (``direct impact accretion''); (2) the 
ejected matter falls back onto star~1 (``self-accretion''); or (3) the 
donor star has completed a full orbital revolution and a new matter 
element is ejected at the periastron of the binary orbit. After the 
accretion of the ejected matter onto star 1 or 2 in cases (1) and (2), 
equations~(\ref{3body}) are integrated as a two-body problem by setting 
$M_3=0$. 

In binaries with eccentric orbits, fallback onto the donor star (case 2) 
is facilitated by the increasing size of the effective Roche Lobe as the 
distance between the stars increases following periastron passage (See 
SWK). The actual outcome of the mass ejection, though, also depends on 
the initial velocity of the ejected matter. In the case where no 
accretion takes place within one orbital revolution (case 3), the system 
evolves as a four-body system until one of the ejected matter elements 
is accreted by star~1 or star~2, or until another mass element is 
ejected at the next periastron passage. The successive ejection of mass 
elements during subsequent periastron passages, possibly causing the 
mass transfer stream to interact with itself, will likely lead to the 
formation of an accretion disk around star~2.  A detailed study of 
accretion disk formation will not be investigated here. We therefore 
restrict ourselves to identifying the regions of parameter space where 
this outcome is expected.

\begin{figure*}
 \vspace{-0.7cm}
\plottwo{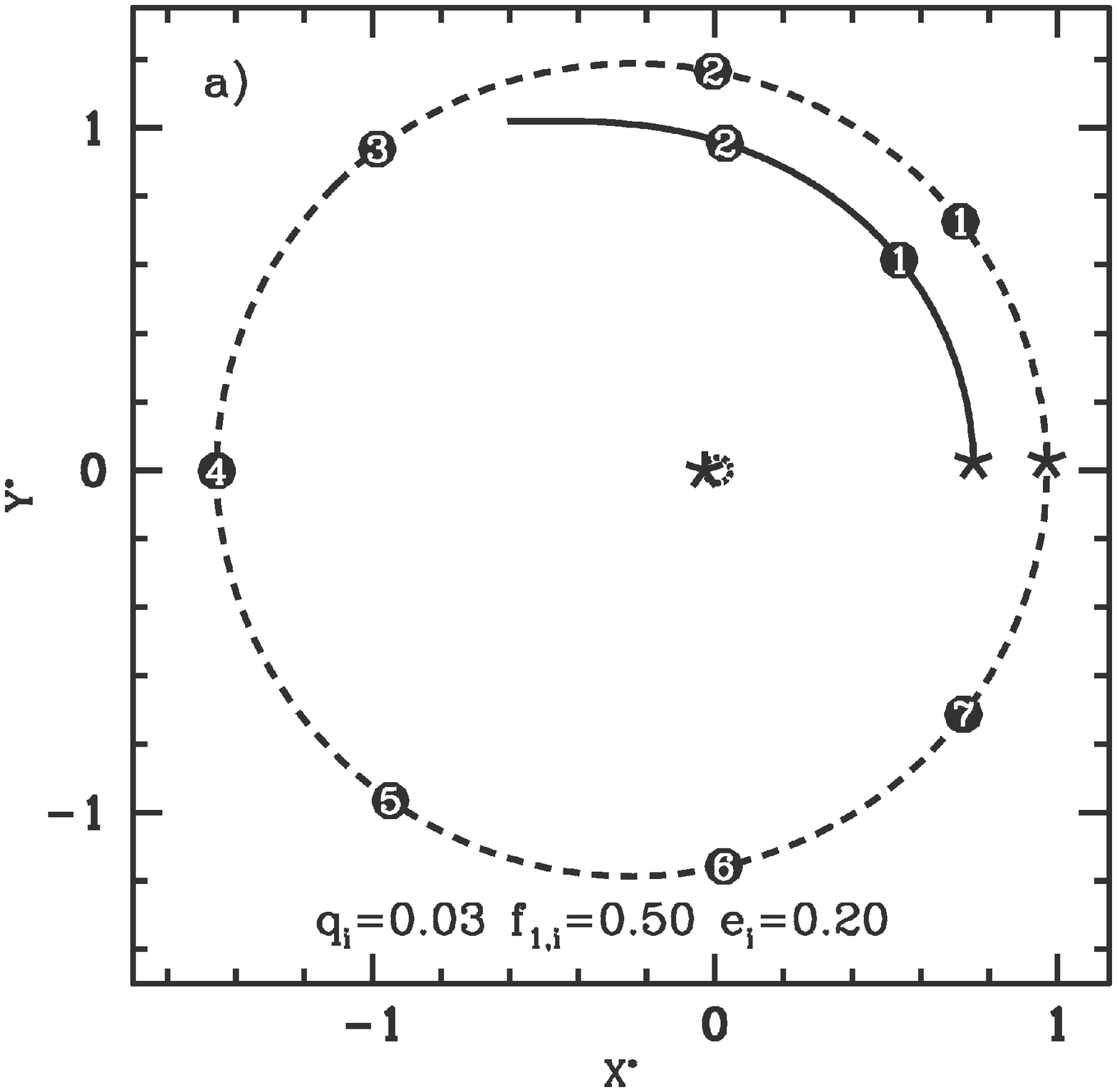}{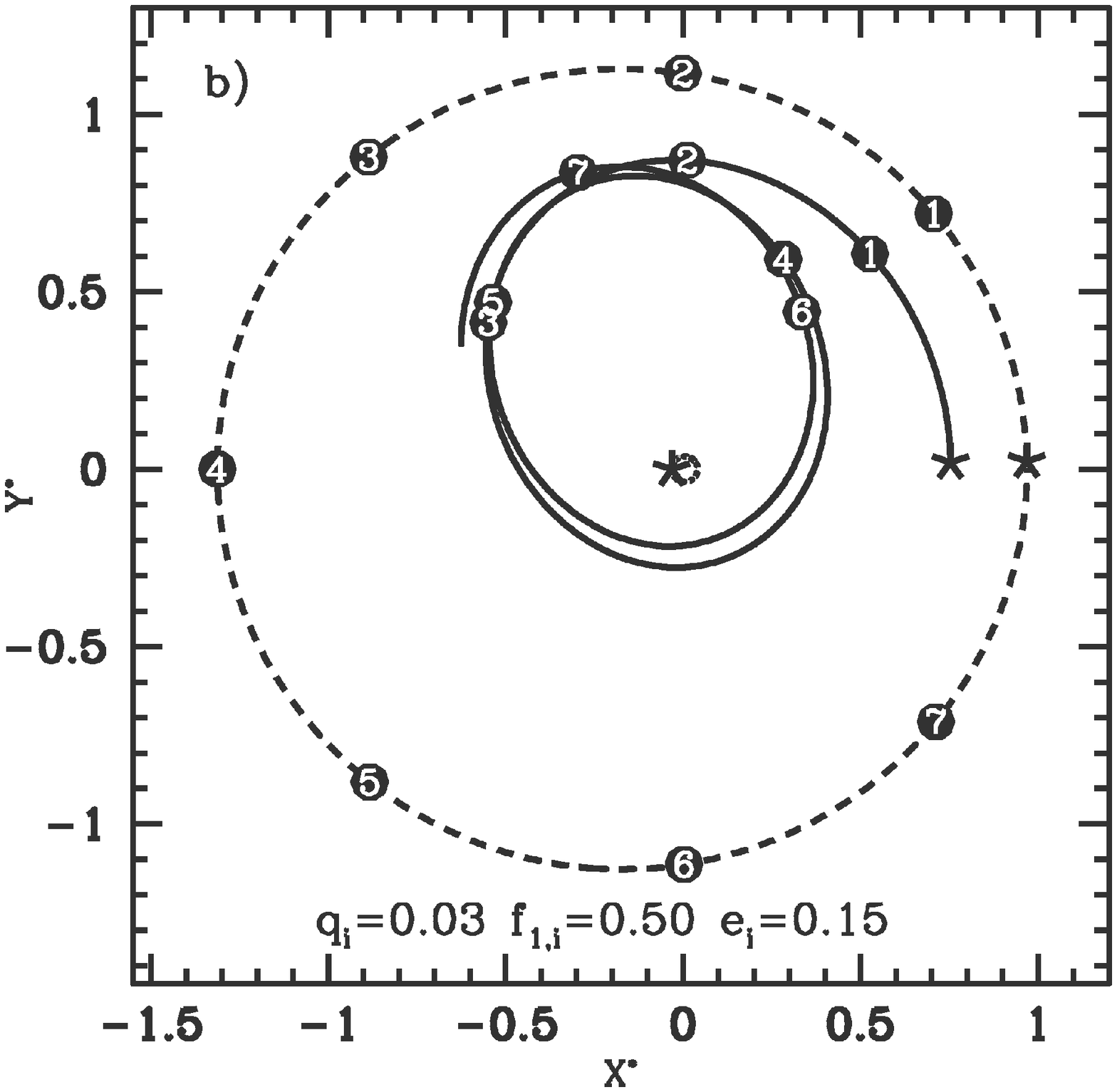}
 \vspace{-0.7cm}
\plottwo{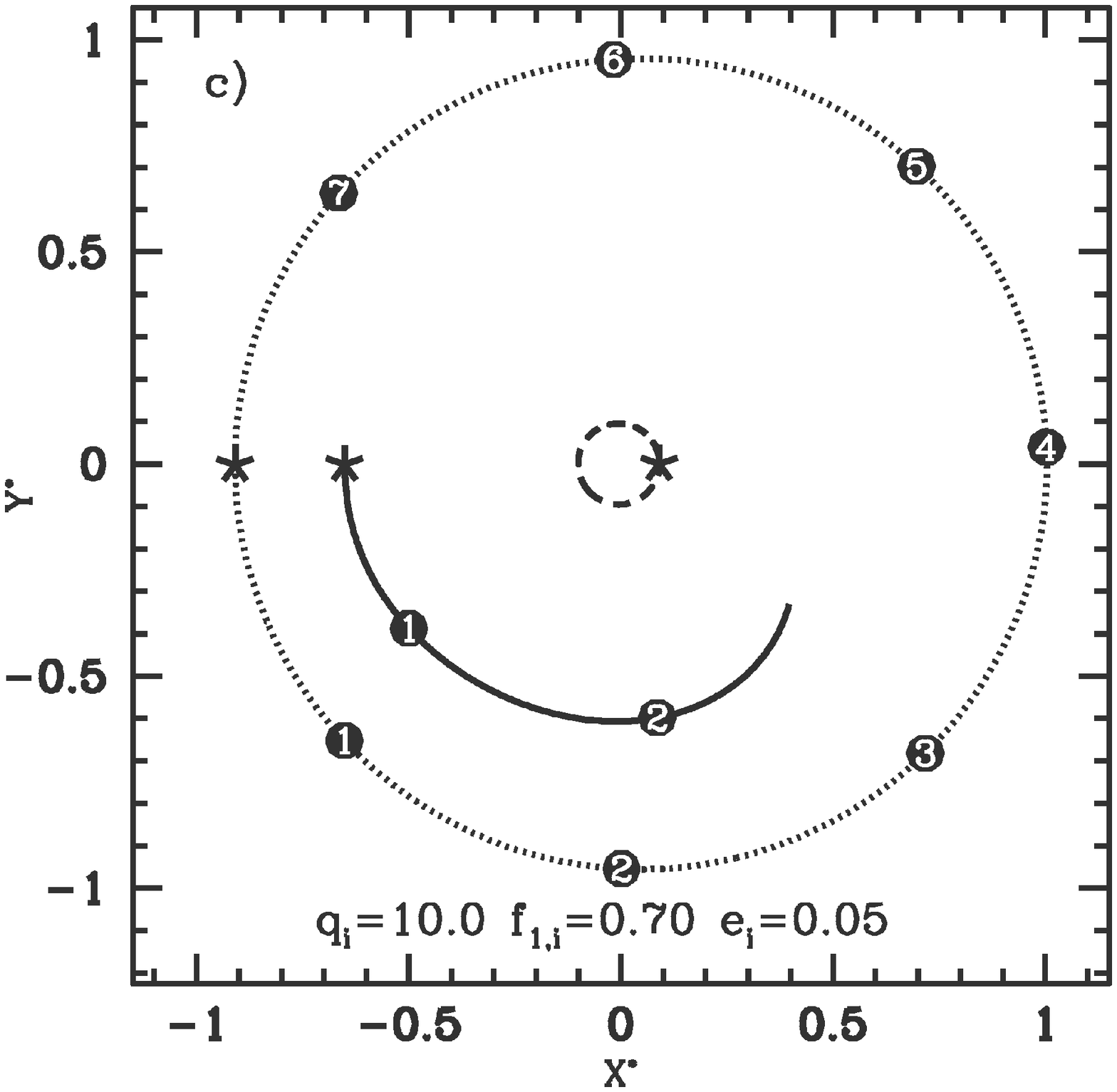}{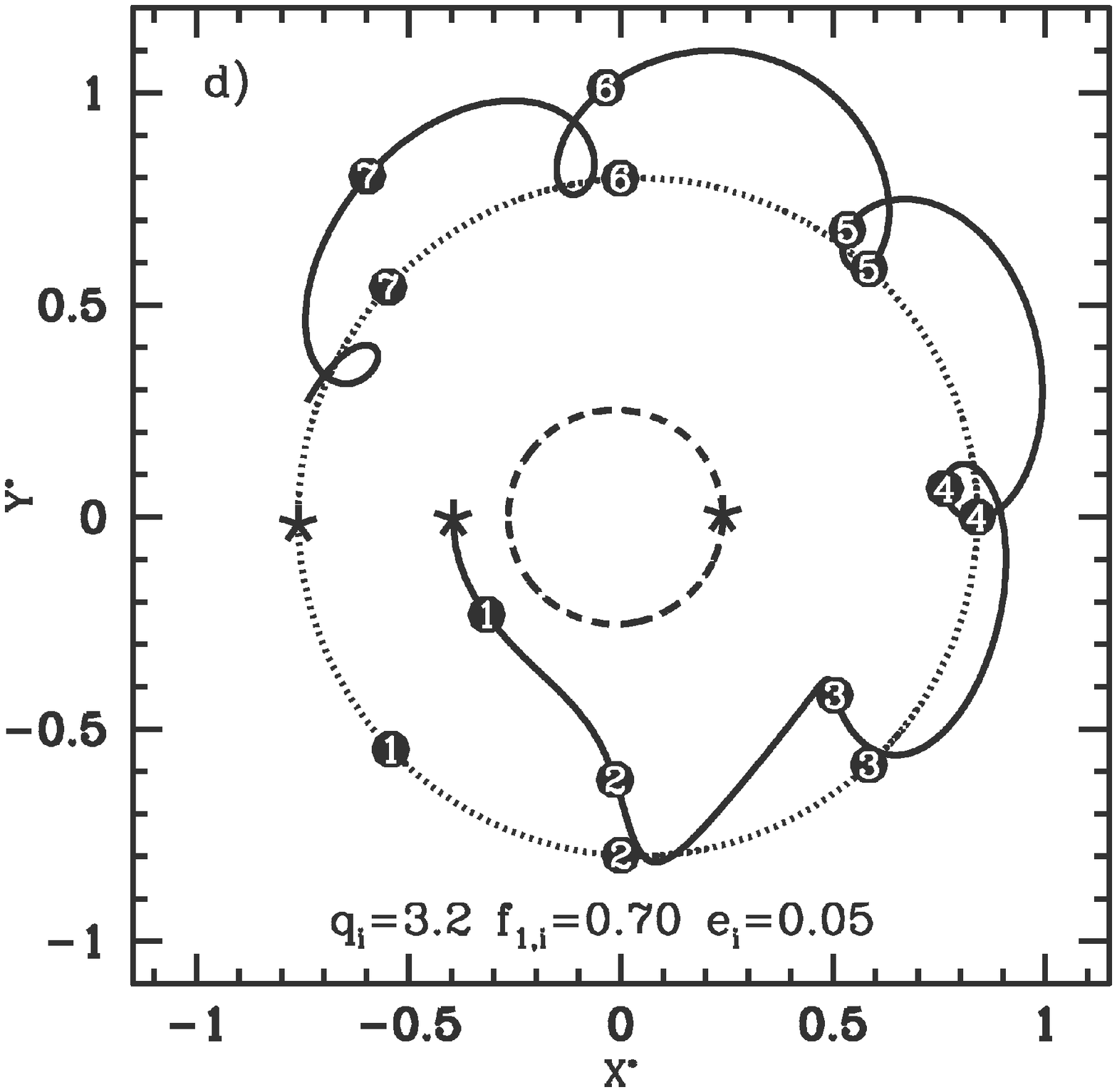}
 \vspace{-0.7cm}
\plottwo{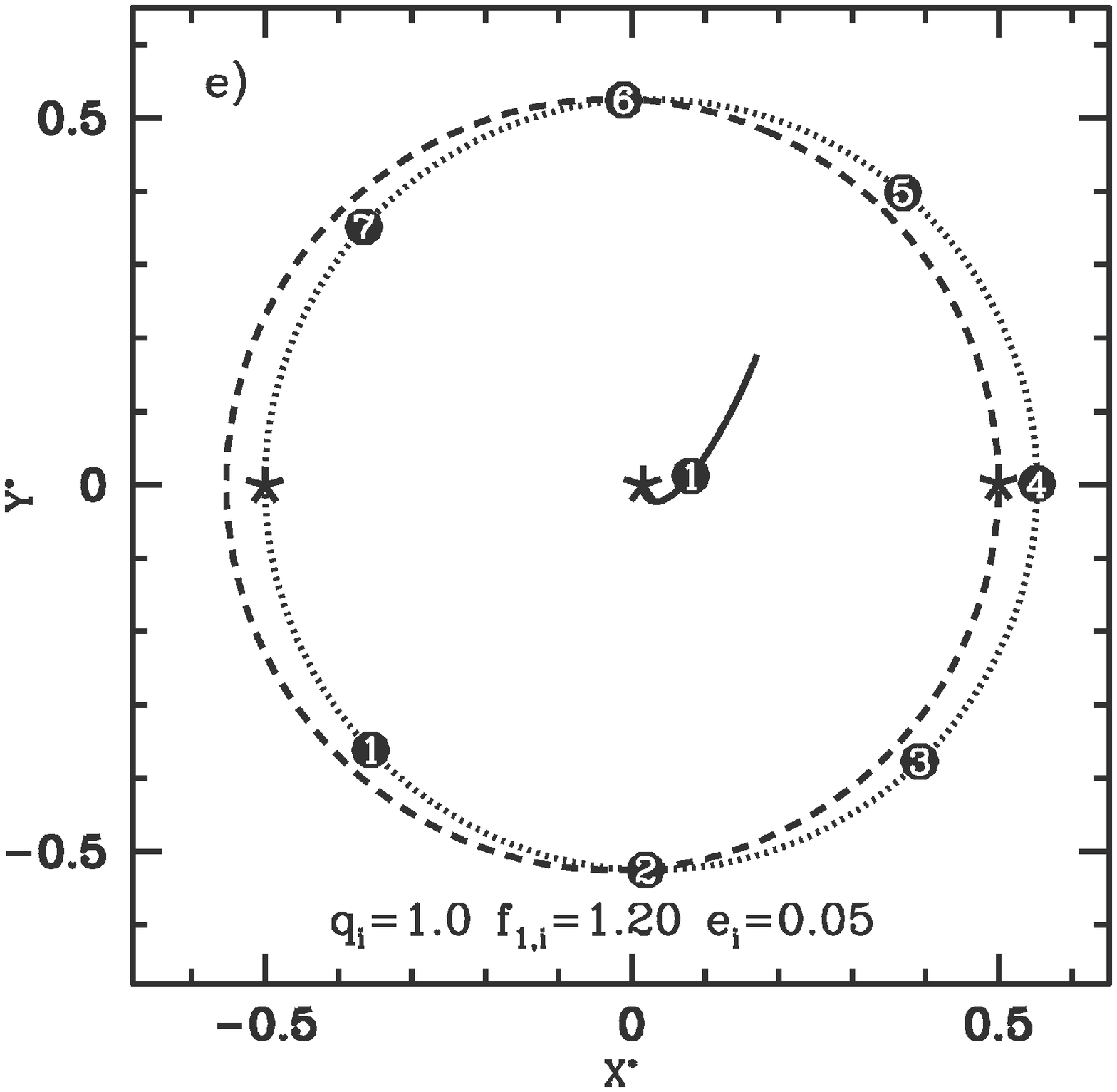}{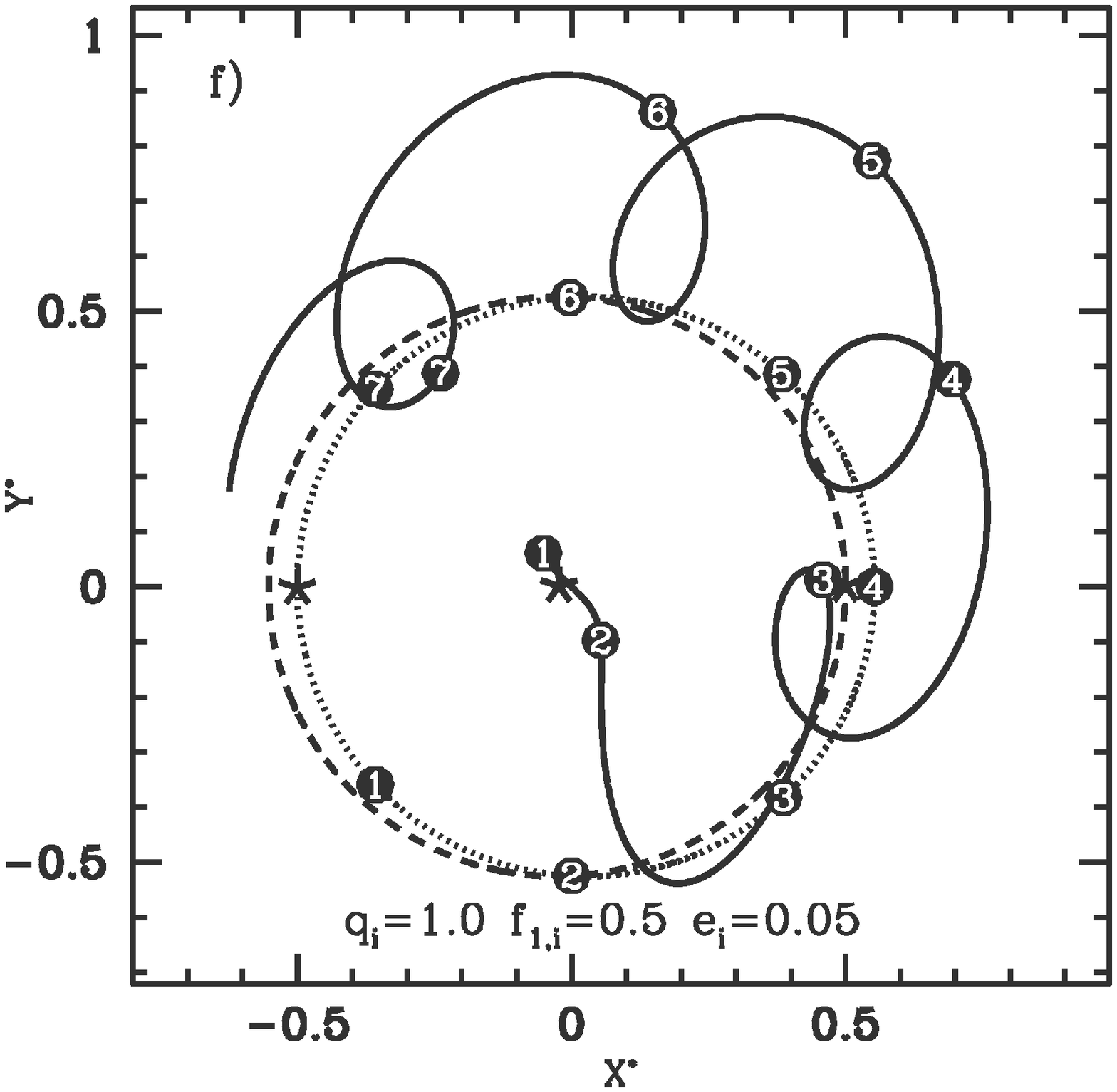}
\caption{\scriptsize Example trajectories of particles ejected from the 
$L_1$ point
in binaries with different initial mass ratios ($q_i=M_{1,i}/M_{2,i}$), 
donor rotation rates ($f_{1,i}$), and orbital eccentricities ($e_i$).  
The $X^*$ and $Y^*$ coordinates are Cartesian coordinates in the orbital 
plane expressed in units of the {\it initial} periastron separation 
between the two stars and with origin at the center of mass of the 
binary system. The solid lines represent the path of the ejected 
particle, the dashed lines the path of the donor star (star~1), and the 
dotted lines the path of the companion star (star~2), which is assumed 
to be a point mass.  The motion of the stars and particles is 
counter-clockwise about the origin, starting from the asterisks 
indicating the initial position of the component stars and the ejected 
matter at the onset of Roche lobe overflow.  The numbered points show 
the position of the stars and the ejected particle at seven different 
equally spaced values of the true anomaly.  The orbits of the stars are 
drawn for one full orbit, regardless of whether the particle has 
accreted or not.  In panels (a), (c), and (e), the particle paths are 
shown up until the point where the particle is accreted back onto the 
donor star.  In panels (b), (d), and (f), the particles undergo neither 
direct impact nor self-accretion. The paths of the particles are 
therefore shown for one complete binary orbit.}
\label{fig-orbits}
\end{figure*}

Some example trajectories of particles ejected from the $L_1$ point of 
star~1 at the periastron of the initial binary orbit are shown in 
Figure~\ref{fig-orbits}, for binaries with different initial mass 
ratios, donor rotation rates, and orbital eccentricities.  The initial 
positions of the stars and the ejected matter are marked by asterisks, 
and the $X^*$ and $Y^*$ coordinates are expressed in units of the 
initial periastron distance.  The motion of the stars is 
counter-clockwise about the origin, and the numbered points on the 
orbits show the position of that object at different equally spaced 
values of the true anomaly.  In all cases the radius of the donor star 
is set equal to the volume-equivalent radius of its effective Roche lobe 
at the periastron of the initial binary orbit (see SWK), and the mass of 
the ejected matter is determined by means of equation~(\ref{M3}). We 
note that as long as $M_3 \ll M_1, M_2$, the ballistic trajectories are 
independent of the actual value of the mass of the ejected matter.  For 
the presented trajectories, star~2 is assumed to be a point mass.

We note that the {\it shape} of the particle trajectories, and 
hence the outcome of the mass overflow, is independent of the size of 
the orbit. This is because the equations of motion and the initial 
positions and velocities are all independent of the size of the 
orbit (see equations [\ref{do1}]-[\ref{do}]).

In panels (a), (c), and (e), the ejected particles fall back onto the 
donor star in less than half an orbit, while in panels (b), (d), and (f) 
the particles end up orbiting the donor's companion. Since the companion 
is assumed to be a point mass with zero radius, none of the displayed 
trajectories lead to direct impact accretion. In panel (d), direct 
impact would occur if the companion had a radius greater than 1.5\% of 
the initial periastron distance. Comparison of panels (a) and (b) shows 
that small changes in the orbital eccentricity can significantly affect 
the type and outcome of the particle trajectories. In this particular 
case, the slightly smaller initial velocity of the ejected particle due 
to the lower eccentricity orbit in panel (b) is enough to cause the 
particle to fall towards star~2 rather than fall back onto star~1. 
Comparison of panels (c) and (d) as well as panels (e) and (f) shows 
that similar transitions in the behavior of the particle trajectories 
can be obtained by changing the initial binary mass ratio or the initial 
donor rotation rate.

\subsection{Spin evolution}

Since the initial velocity of the matter ejected at periastron depends 
on the rotation of the donor star, it is important to keep track of 
changes in the star's rotational angular velocity due to mass loss and 
accretion. To calculate the change in the rotational angular velocity of 
the donor star due to the loss of an element of mass $M_3$, we note 
that the total angular momentum $\vec{\cal{L}}_I$ in the system right 
before the ejection of matter at periastron is given by
\begin{eqnarray} 
\vec{\cal{L}}_I &=&
I_{1,I}\vec{\Omega}_{1,I} + I_{2,I}\vec{\Omega}_{2,I} \nonumber \\ && +
M_{1,I}\vec{R}_{1,I}\times\vec{V}_{1,I} +
M_{2,I}\vec{R}_{2,I}\times\vec{V}_{2,I}, 
\end{eqnarray} 
where $I_1$ and $I_2$ are the moments of inertia, $\vec{R}_1$ and $\vec{R}_2$
the position vectors, and $\vec{V}_1$ and $\vec{V}_2$ the velocities of
star~1 and star~2, respectively, measured with respect to the inertial 
frame of reference. The subscript ``I'' denotes that the quantities are 
to be considered right before the ejection of matter at periastron. 
Similarly, the total angular momentum $\vec{\cal{L}}_F$ in the system 
right after the ejection of matter at periastron is given by
\begin{eqnarray} 
\vec{\cal{L}}_F &=& I_{1,F}\vec{\Omega}_{1,F} +
I_{2,F}\vec{\Omega}_{2,F} \nonumber \\ && +
M_{1,F}\vec{R}_{1,F}\times\vec{V}_{1,F} +
M_{2,F}\vec{R}_{2,F}\times\vec{V}_{2,F} \nonumber \\ && +
M_3\vec{R}_3\times\vec{V}_3, 
\end{eqnarray} 
where $\vec{R}_3$ and $\vec{V}_3$ are the position and velocity vectors
of the ejected mass element at the time of ejection measured with 
respect to the inertial frame of reference. The subscript ``F'' denotes 
that the quantities are to be considered right after the ejection of 
matter at periastron.  For the uniform density spheres considered here, 
the moment of inertia is $I=0.4M{\cal R}^2$.  For a more centrally 
condensed object the moment of inertia will decrease, reducing the 
contribution of the spin to the total angular momentum.

Since the ejection of matter by star~1 does not instantaneously affect
the companion star, all quantities pertaining to star~2 are identical 
right before and right after the instantaneous ejection of a mass 
element.  Conservation of total angular momentum therefore yields
\begin{eqnarray}
\label{eq-rotrate1}
\vec{\Omega}_{1,F} &=& \frac{1}{I_{I,F}}\left( I_{1,I}\vec{\Omega}_{1,I} 
+ M_{1,I}\vec{R}_{1,I}\times\vec{V}_{1,I} \right. \nonumber \\
&&\left. - M_{1,F}\vec{R}_{1,F}\times\vec{V}_{1,F} - 
M_3\vec{R}_3\times\vec{V}_3 \right).  \label{o1f}
\end{eqnarray}
In this equation, all quantities on the right-hand side are readily 
obtained from the initial conditions and the adopted mass overflow rate, 
except for $\vec{R}_{1,F}$ and $\vec{V}_{1,F}$. To determine 
$\vec{R}_{1,F}$ we assume $M_3$ is ejected perfectly inelastically, 
i.e., with no loss of linear momentum\footnote{We note that, in a 
perfectly inelastic collision energy is not conserved.  We assume the 
existence of some mechanism, either internal or external, which allows 
the ejected particle to become physically separate from the donor star.  
Similarly for accretion, some process must exist to allow the particle 
and accretor to physically merge.  In either case, the change in the 
total energy of the system is likely to happen at the expense of the 
thermal energy of the component stars.  For further discussion of the 
change in the total energy, see \S\ref{sec-sequences}}.  Furthermore, 
since the ejection is instantaneous, the center of mass of star~1 before 
the ejection is identical to the center of mass of the system consisting 
of the particle and star~1 after the ejection.	Thus, the final position 
of star~1 is given by
\begin{equation}
\vec{R}_{1,F} = \frac{1}{M_{1,F}}\left( M_{1,I} \vec{R}_{1,I} - 
M_3\vec{R}_3 \right).  \label{r1f}
\end{equation}
The velocity $\vec{V}_{1,F}$ is determined from the conservation of 
linear momentum:
\begin{equation}
\vec{V}_{1,F} = \frac{1}{M_{1,F}}\left( 
M_{1,I}\vec{V}_{1,I}-M_3\vec{V}_3 \right).  \label{v1f}
\end{equation}

To calculate the change in the rotational angular velocity due to the 
fallback of the ejected matter onto the donor star, it suffices to 
replace $M_3$ by $- M_3$, and $\vec{R}_3$ and $\vec{V}_3$ by the 
position and velocity vector of the ejected mass element at the time of 
accretion in equations~(\ref{o1f})--(\ref{v1f}).

The change in the rotational angular velocity of star~2 due to direct 
impact accretion is derived similarly:
\begin{eqnarray}
\label{eq-rotrateDI}
\vec{\Omega}_{2,F} &=& \frac{1}{I_{2,F}} \left[ 
I_{2,I}\vec{\Omega}_{2,I} 
+ M_{2,I}\vec{R}_{2,I}\times\vec{V}_{2,I} \right. \nonumber \\
&&\left. + M_3\vec{R}_3\times\vec{V}_3 - 
M_{2,F}\vec{R}_{2,F}\times\vec{V}_{2,F} \right]
\end{eqnarray}
with
\begin{equation}
\vec{R}_{2,F} = \frac{1}{M_{2,F}}\left( M_{2,I}\vec{R}_{2,I} + 
M_3\vec{R}_3 \right),
\end{equation}
and
\begin{equation}
\label{eq-V2F}
\vec{V}_{2,F} = \frac{1}{M_{2,F}}\left( M_{2,I}\vec{V}_{2,I} + 
M_3\vec{V}_3 \right).
\end{equation}
Here, the subscripts ``I'' and ``F'' denote the quantities are to be 
considered right before and right after the accretion of matter, 
respectively, and $\vec{R}_3$ and $\vec{V}_3$ are the position and 
velocity vector of the ejected mass element at the time of accretion.

\subsection{Calculation of the Orbital Elements}

Due to the gravitational interactions between the ejected matter and the 
component stars of the binary system, the orbital motion of the stars is 
no longer purely Keplerian. In particular, the semi-major axis and 
eccentricity are not constant in time. We calculate the instantaneous 
value of the osculating orbital elements of the binary system at each 
timestep using \citep{1960aitc.book.....S}
\begin{eqnarray}
\label{eq-calce}
e^2 &=& 1- \frac{ \left| \vec{r}_{\rm rel} \times
\vec{v}_{\rm rel} \right|^2}{G\left( M_1+M_2\right) a}, \\ \frac{1}{a}
&=& \frac{2}{|\vec{r}_{\rm rel}|} - \frac{v^2_{\rm rel}}{G\left(
M_1+M_2\right)},
\end{eqnarray}
where $\vec{r}_{\rm rel}$ and $\vec{v}_{\rm rel}$ are, respectively, the
position and velocity vectors of star~2 with respect to star~1. 
The osculating orbital elements represent the value of $a$ 
and $e$ that the orbit would have if the perturbative forces were 
removed at that instant.  To keep track of changes in the position of 
the periastron due to mass overflow, we also calculate the instantaneous 
true anomaly of the binary system from
\begin{equation}
\label{eq-nu}
\cos{\nu} = \frac{1}{e}\left[ \frac{a\left( 1-e^2 \right)}{|\vec{r}_{\rm 
rel}|} -1\right].
\end{equation}

\section{Parameter Space for Direct Impact and Self-Accretion}
\label{sec-parameterspace}

In order to investigate the parameter space leading to different 
outcomes of mass overflow (direct impact accretion, self-accretion, or 
possible disk formation) in binaries with eccentric orbits, we performed 
an extensive set of ballistic orbit calculations for binaries with 
different initial mass ratios $q_i$, initial donor rotation rates 
$f_{1,i}$, and initial orbital eccentricities $e_i$ at the onset of mass 
overflow. As in \S\ref{sec-trajectories}, we set the donor's radius 
equal to the volume equivalent radius of its effective Roche lobe at the 
periastron of the initial binary orbit. We recall that as long 
as $M_3 \ll M_1, M_2$ the problem is fully determined by $q_i$, 
$f_{1,i}$, and $e_i$, and is independent of the actual values of the 
mass overflow rate $\dot{M}_0$, donor mass $M_1$, donor radius, ${\cal 
R}_1$, and initial orbital semi-major axis $a_i$. Star~2 is again 
treated as a point mass.

\begin{figure}
\plotone{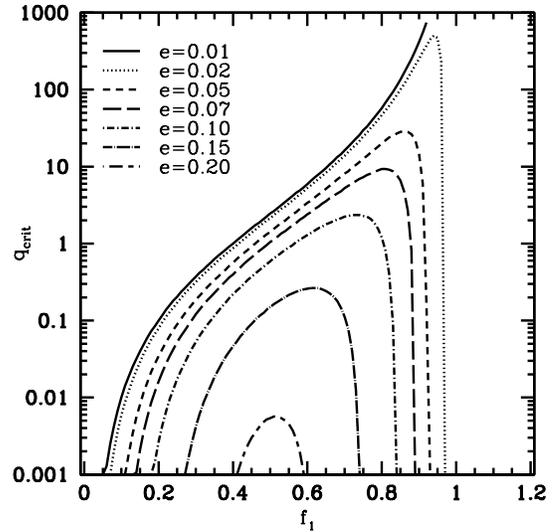}
\caption{The critical mass ratio $q_{\rm crit}$ above which systems 
undergo self-accretion as a function of the donor's initial rotation 
rate $f_{1,i}$, for initial orbital eccentricities ranging from 0.01 to 
0.2.  For $q_i>q_{\rm crit}$, systems undergo self-accretion, while for 
$q_i<q_{\rm crit}$ systems undergo direct impact accretion or form an 
accretion disk, depending on the radius of the accretor.  We note that 
for low initial eccentricities, $e_i \lesssim 0.05$, the delta-function 
approximation of the mass overflow rate given by equation~(\ref{eq-m0}) 
likely breaks down, so that at low eccentricities disk formation or 
direct impact may occur for a larger region of parameter space than 
presented here.}
\label{fig-qcrit}
\end{figure}

For each initial orbital eccentricity and donor rotation rate,
we find a critical mass ratio, $q_{\rm crit}$, such that for $q_i > 
q_{\rm crit}$ the system undergoes self-accretion, while for $q_i < 
q_{\rm crit}$, the system may undergo direct impact accretion or form an
accretion disk, depending on the radius of star~2. The critical mass
ratio is shown in Figure~\ref{fig-qcrit} as a function of the donor's
initial rotation rate for initial orbital eccentricities ranging from
0.01 to 0.2. In this eccentricity range, $q_{\rm crit}$ decreases with
increasing initial orbital eccentricity, increasing the parameter space
leading to self-accretion.

For initial eccentricities $e_i > 0.2$, mass overflow always leads to 
self-accretion if the initial binary mass ratio $q_i > 0.006$.  In 
high-eccentricity systems direct impact accretion therefore only occurs 
for very low mass ratios. For initial donor rotation rates in the range 
$1.0 \le f_{1,i} \le 1.2$, mass ejection at periastron results in 
self-accretion for all initial mass ratios $10^{-3} < q_i < 10^3$, 
independent of the initial orbital eccentricity. For the purpose of this 
paper, we did not thoroughly explore the parameter space for $f_{1,i} > 
1.2$, though calculation of some sample trajectories of mass ejected 
from highly super-synchronously rotating donor stars indicates the 
matter may start orbiting the donor star before leading to 
self-accretion.

When $q_i < q_{\rm crit}$, depending on the accretor's radius ${\cal 
R}_2$ the ejected matter can directly impact the companion star or go 
into orbit around it, possibly forming an accretion disk. While studying 
disk formation is beyond the capabilities of our ballistic trajectory 
code, we can investigate when direct impact rather than disk formation 
is expected to occur based on the ballistic trajectories of the ejected 
mass elements. For this purpose, we treat star~2 as a point mass and 
determine the minimum distance $r_{\rm min}$ between the ejected matter 
and star~2 during a single orbit for ballistic trajectories not leading 
to self-accretion. Hence, if ${\cal R}_2 > r_{\rm min}$, direct impact 
occurs within one binary orbital period.  If ${\cal R}_2 < r_{\rm min}$, 
the ejected matter may still impact star~2 after one orbital period, but 
it may also orbit star~2 until it interacts with other ejected mass 
elements and forms an accretion disk.

\begin{figure}
\plotone{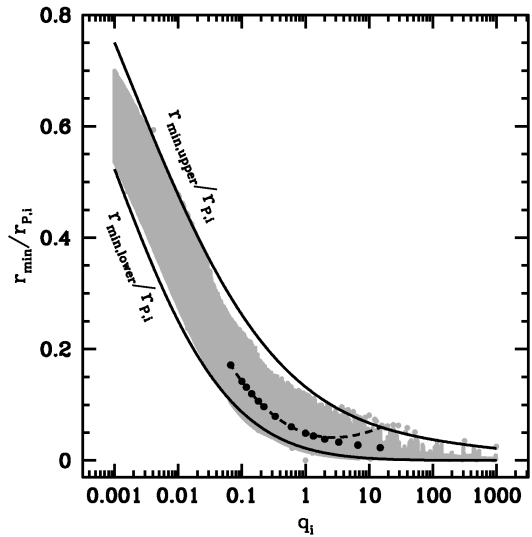}
\caption{The minimum distance $r_{\rm min}/r_{P,i}$ from star~2 (assumed 
to be a point mass) reached by matter ejected from the $L_1$ point of 
the donor star at the periastron of the binary orbit as a function of 
the logarithm of the initial binary mass ratio $q_i$. The distance is 
measured in the orbital plane and is normalized to the initial 
periastron separation $r_{P,i}$ between the two stars.  For a given mass 
ratio, the grey dots represent minimum distances for different initial 
donor rotation rates $f_{1,i}$ ranging from 0.01 to 1.0, and initial 
orbital eccentricities $e_i$ ranging from 0.01 to 0.2. The solid lines 
represent the approximate upper and lower limits on the minimum distance 
as a function of $q_i$ given by equations~(\ref{eq-rlower}) and 
(\ref{eq-rupper}).  The black circles are the data of 
\citet{1975ApJ...198..383L} for the closest approach of a ballistic 
particle in a circular, synchronous binary system, and the dashed line 
is the fit to the data of \citet{1975ApJ...198..383L} given by 
\citet{2001A&A...368..939N}. Both the fit and the data shown are 
adjusted to account for the different definitions of the mass ratio 
given in those papers.}
\label{fig-rmin}
\end{figure}

The minimum distance $r_{\rm min}/r_{P,i}$ is shown in 
Figure~\ref{fig-rmin} as a function of the logarithm of the initial 
binary mass ratio $q_i$. For each initial mass ratio, the grey dots 
represent the minimum distance associated with different initial donor 
rotation rates ($0.01 \le f_{1,i} \le 1.0$) and initial orbital 
eccentricities ($0.01 \le e_i \le 0.2$).  The main trend of $r_{\rm 
min}/r_{P,i}$ as a function of $q_i$ is to decrease with increasing 
values of $q_i$. This is partly because the Roche lobe of star~2 is 
smaller for large mass ratios than for small mass ratios, or, 
equivalently, the initial position of the ejected matter, the $L_1$ 
point of star~1, is relatively closer to star~2 for large mass ratios 
than for small mass ratios (see SWK).

The lower and upper edges of the grey dotted region in
Figure~\ref{fig-rmin} can be approximated by
\begin{equation}
\frac{r_{\rm min, lower}}{r_{P,i}} = \exp \left[ 
{-0.16(\log_{10}{q_i}+4.8)^2 - 0.13} 
\right],
\label{eq-rlower}
\end{equation}
and 
\begin{eqnarray}
\frac{r_{\rm min, upper}}{r_{P,i}} &=& \exp \left[ 
{-0.1(\log_{10}{q_i}+4.8)^2} \right] \nonumber \\
&-& 0.01\log_{10}{q_i} + 0.05.
\label{eq-rupper}
\end{eqnarray}
Hence, if ${\cal R}_2 > r_{\rm min, upper}$, the system always 
undergoes direct impact accretion within one orbital period. If ${\cal 
R}_2 < r_{\rm min, lower}$, the ejected matter always avoids 
impacting the donor's companion star within one orbital period, so that 
an accretion disk may be formed. If $r_{\rm min, lower} < {\cal 
R}_2 < r_{\rm min, upper}$, the outcome of the mass overflow depends on 
the initial donor rotation rate and orbital eccentricity.

For comparison, the black circles of Figure~\ref{fig-rmin} show the 
distance of closest approach for a particle in a circular, synchronous 
binary system as calculated by \citet{1975ApJ...198..383L} (their 
$\varpi_{\rm min}$), where we have taken into account the difference 
between their and our definition of the mass ratio.  We note that this 
data is consistent with our calculation of the distance of closest 
approach, and that the inclusion of non-synchronous rotation and/or 
eccentricity can either increase or decrease the distance of closest 
approach.  The dashed line through these data points is the fit to the 
data of \citet{1975ApJ...198..383L} given by 
\citet{2001A&A...368..939N}.  The fit is very accurate for $q_i \la 2$, 
but should not be extrapolated to large $q_i$.  This does not affect the 
results of \citet{2001A&A...368..939N}, as they are interested in double 
white dwarf systems where the mass ratio of the system is generally less 
than unity.

\section{Orbital Evolution Due to Mass Overflow}
\label{sec-orbitevolution}

Using the framework described in \S\ref{sec-MTinecc}, we can now 
investigate the secular evolution of the orbital elements due to 
the effects of mass overflow in eccentric binary systems. Combining
equations~(\ref{eq-rotrate1})--(\ref{eq-V2F}) with 
equations~(\ref{eq-calce})--(\ref{eq-nu}), it can be seen that the 
changes in the orbital elements, and thus the time scales of orbital 
evolution, depend on the binary component masses, the donor radius 
and rotation rate (through the initial particle velocity), and the 
orbital elements (also through the initial particle velocity).  
The outcome of the mass overflow (e.g. whether or not direct
impact accretion occurs) also depends on the radius of the accretor.

Once mass overflow starts, the radius of the donor star evolves due to 
the combined effects of mass loss and stellar evolution. In addition, 
the radius of the donor's effective Roche lobe changes due the evolution 
of its rotational angular velocity, the orbital elements, and the binary 
mass ratio (SWK). In order for mass overflow in eccentric binaries to be 
stable and long-lived, the radial evolution of the donor star must 
follow the evolution of its effective Roche lobe at the periastron of 
the binary orbit. In this section, we explore the potential long-term 
evolution of effective Roche lobe overflowing binaries with eccentric 
orbits, {\it assuming} that after the onset of mass overflow the donor 
star's radius remains equal to the radius of its effective Roche lobe at 
each subsequent periastron passage. Whether the response of the star to 
mass loss renders this possible depends on its internal structure and on 
the binary properties. Since we here treat the stars as rigid spheres, 
we defer questions on the longevity and the stability of mass overflow 
to future investigations.

To illustrate the effects of mass overflow in eccentric binaries, we 
consider a binary star system consisting of two rigid spheres with radii 
representative of main-sequence stars. The first rigid sphere (``star 
1'') is assumed to have a radius representative of a star near the end 
of core-hydrogen burning when the main-sequence radius is maximal, while 
the second rigid sphere (``star~2'') is assumed to be a zero-age 
main-sequence star. Such a binary consisting of two non-coeval component 
stars can be formed through dynamical interactions in globular clusters. 
The adopted configuration allows us to illustrate the orbital evolution 
time scales of binaries with initial mass ratios $q_i<1$ as well as 
binaries with initial mass ratios $q_i>1$. In addition, a zero-age 
main-sequence accretor constitutes a big enough ``target'' to provide a 
non-negligible parameter space leading to direct impact accretion. A 
study of direct impact accretion in compact object binaries, double 
white dwarfs in particular, will be presented in a forthcoming paper.

\subsection{Orbital Evolution Timescales}
\label{sec-orbitev}

\begin{figure*}
\plotone{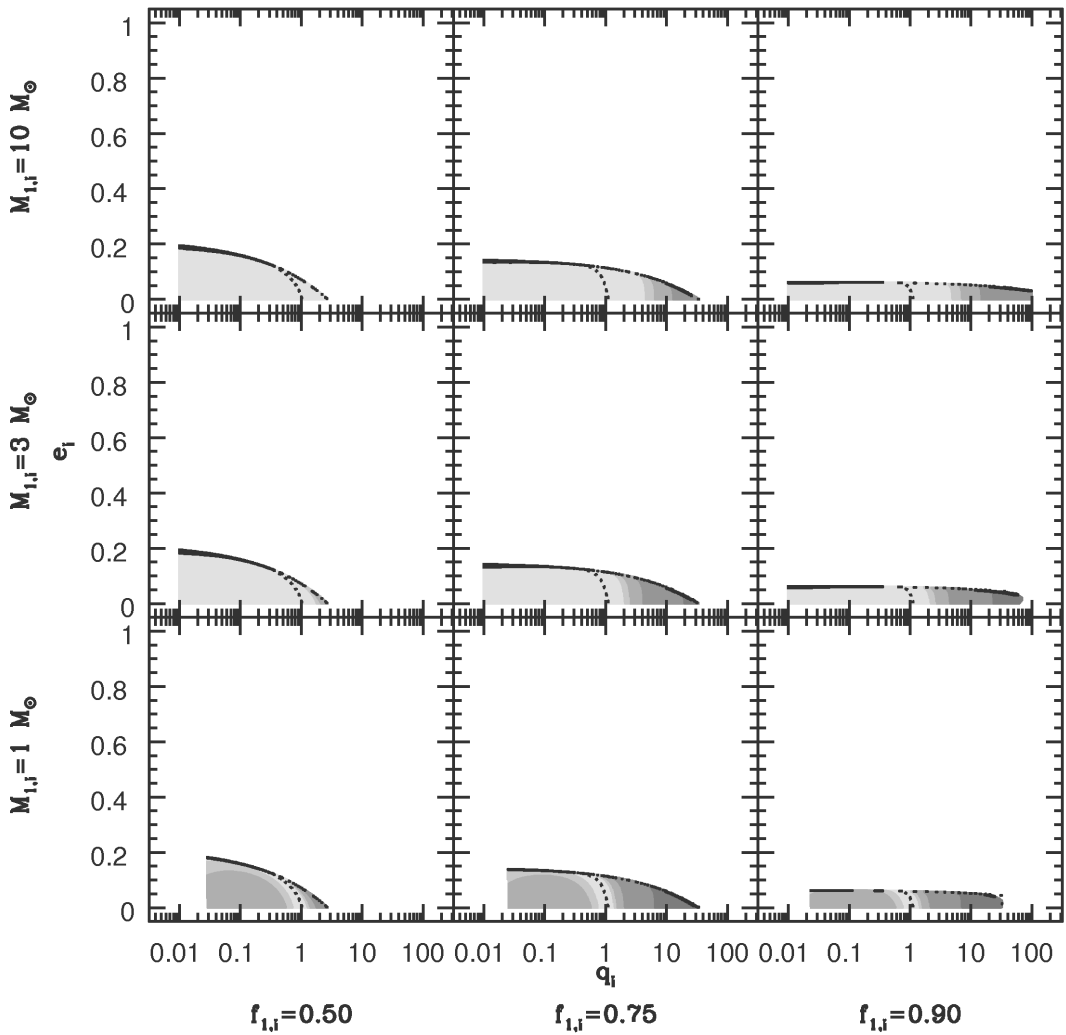}
\caption{Timescales for the evolution of the semi-major axis $a$ as a 
function of the initial mass ratio $q_i$ and initial orbital 
eccentricity $e_i$ in systems undergoing direct impact accretion. The 
donor star is assumed to be near the end of the main sequence and to 
have an initial mass $M_{1,i}=1, 3$, or $10\,M_\sun$, and an initial 
rotation rate $f_{1,i}= 0.5, 0.75$, or $0.9$. The accretor is assumed to 
be a zero-age main-sequence star with a mass determined by the initial 
mass ratio $q_i$. The mass overflow rate $\dot{M}_0$ at periastron is 
$-10^{-9}\,M_\odot\, {\rm yr^{-1}}$. The different shades of gray 
represent regions in the $(q_i,e_i)$ parameter space with different 
characteristic timescales of orbital evolution. From the darkest to the 
lightest shade of gray, the timescales are: $0\, {\rm Gyr}<|\tau_a|<1\, 
{\rm Gyr}$, $1\, {\rm Gyr}<|\tau_a|<5\, {\rm Gyr}$, $5\, {\rm 
Gyr}<|\tau_a|<10\, {\rm Gyr}$, $10\, {\rm Gyr}<|\tau_a|<15\, {\rm Gyr}$, 
$15\, {\rm Gyr}<|\tau_a|$. The dotted black line near $q_i=1$ divides 
the regions of the $(q_i,e_i)$ parameter space where the semi-major axis 
increases as a function of time (to the left of the line) from the 
regions of the $(q_i,e_i)$ parameter space where the semi-major axis 
decreases as a function of time (to the right of the line). The black 
shaded area corresponds to regions of the parameter space where the mode 
of mass overflow switches from direct impact to self-accretion or 
possible disk formation within the first 100 orbits after the onset of 
mass overflow.  This area occupies only a small space between the the 
direct impact and self-accretion regimes. Unshaded regions correspond to 
regions of the parameter space where mass overflow either does not lead 
to direct impact accretion or the zero-age main-sequence accretor is so 
large that a contact binary rather than a semi-detached binary is 
formed.}
\label{fig-dadtDI}
\end{figure*}

\begin{figure*}
\plotone{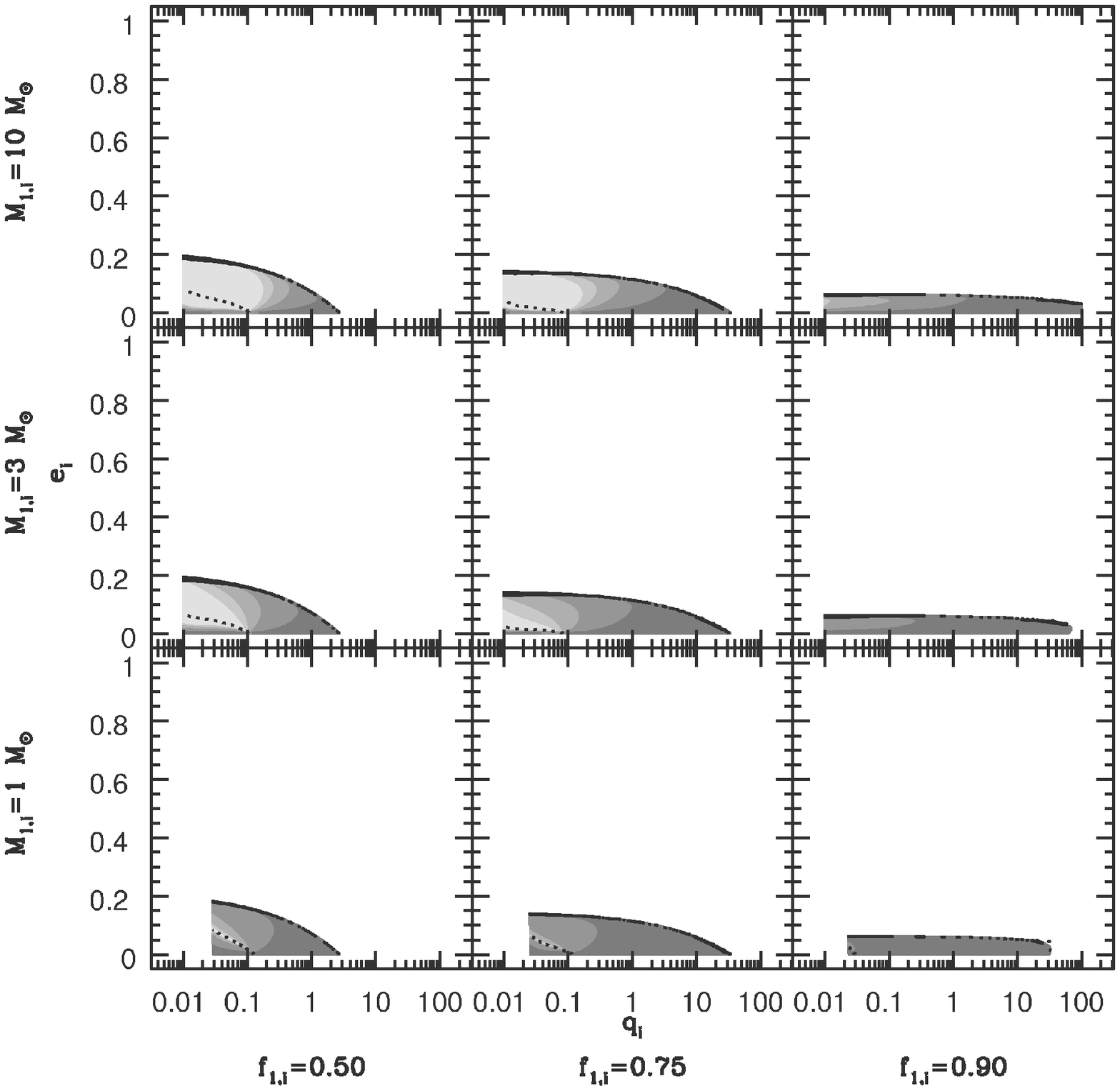}
\caption{Same as Figure~\ref{fig-dadtDI}, but for the evolution 
of the orbital eccentricity $e$.}
\label{fig-dedtDI}
\end{figure*}

\begin{figure*}
\plotone{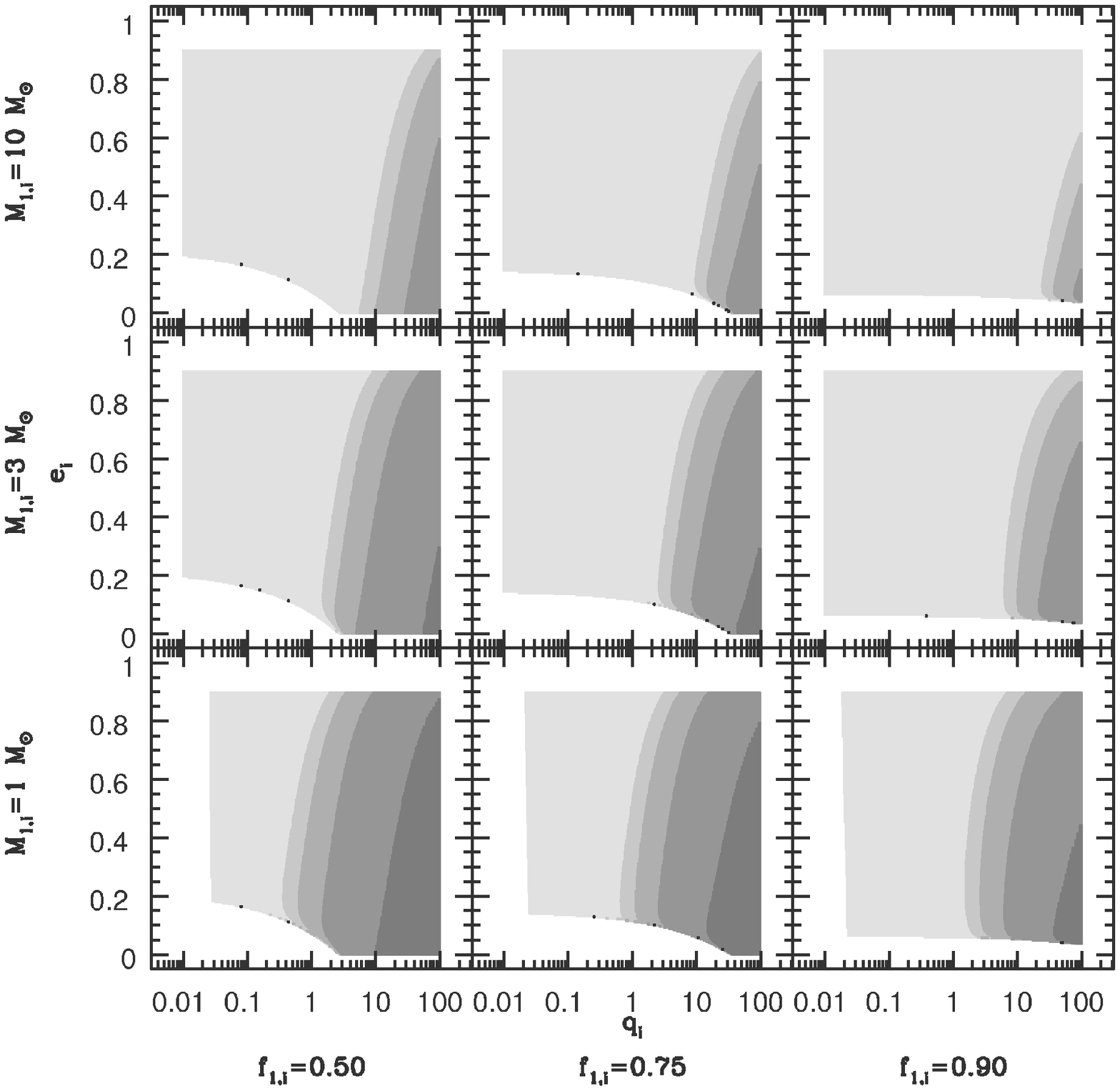}
\caption{Timescales for the evolution of the semi-major axis $a$ as a 
function of the initial mass ratio $q_i$ and initial orbital 
eccentricity $e_i$ in systems undergoing self-accretion. The donor star 
is assumed to be near the end of the main sequence and to have an 
initial mass $M_{1,i}=1, 3$, or $10\,M_\sun$, and an initial rotation 
rate $f_{1,i}= 0.5, 0.75$, or $0.9$. The binary companion is assumed to 
be a zero-age main-sequence star with a mass determined by the mass 
ratio $q_i$. The mass overflow rate at periastron is $\dot{M}_0 = 
-10^{-9}\,M_\odot\, {\rm yr^{-1}}$. The different shades of gray 
represent regions in the $(q_i,e_i)$ parameter space with different 
characteristic timescales of orbital evolution. From the darkest to the 
lightest shade of gray, the timescales are: $0\, {\rm Gyr}<|\tau_a|<1\, 
{\rm Gyr}$, $1\, {\rm Gyr}<|\tau_a|<5\, {\rm Gyr}$, $5\, {\rm 
Gyr}<|\tau_a|<10\, {\rm Gyr}$, $10\, {\rm Gyr}<|\tau_a|<15\, {\rm Gyr}$, 
$15\, {\rm Gyr}<|\tau_a|$. The timescales always represent a decrease of 
the semi-major axis. The black points at low eccentricity correspond to 
regions of the parameter space where the mode of mass overflow switches 
from self-accretion to direct impact accretion or possible disk 
formation within the first 100 orbits after the onset of mass overflow.  
These points occupy only a small space between the the direct impact and 
self-accretion regimes. Unshaded regions correspond to regions of the 
parameter space where either mass overflow does not lead to 
self-accretion or the zero-age main-sequence accretor is so large that a 
contact binary rather than a semi-detached binary is formed.}
 \label{fig-dadtSA} 
\end{figure*}

\begin{figure*}
\plotone{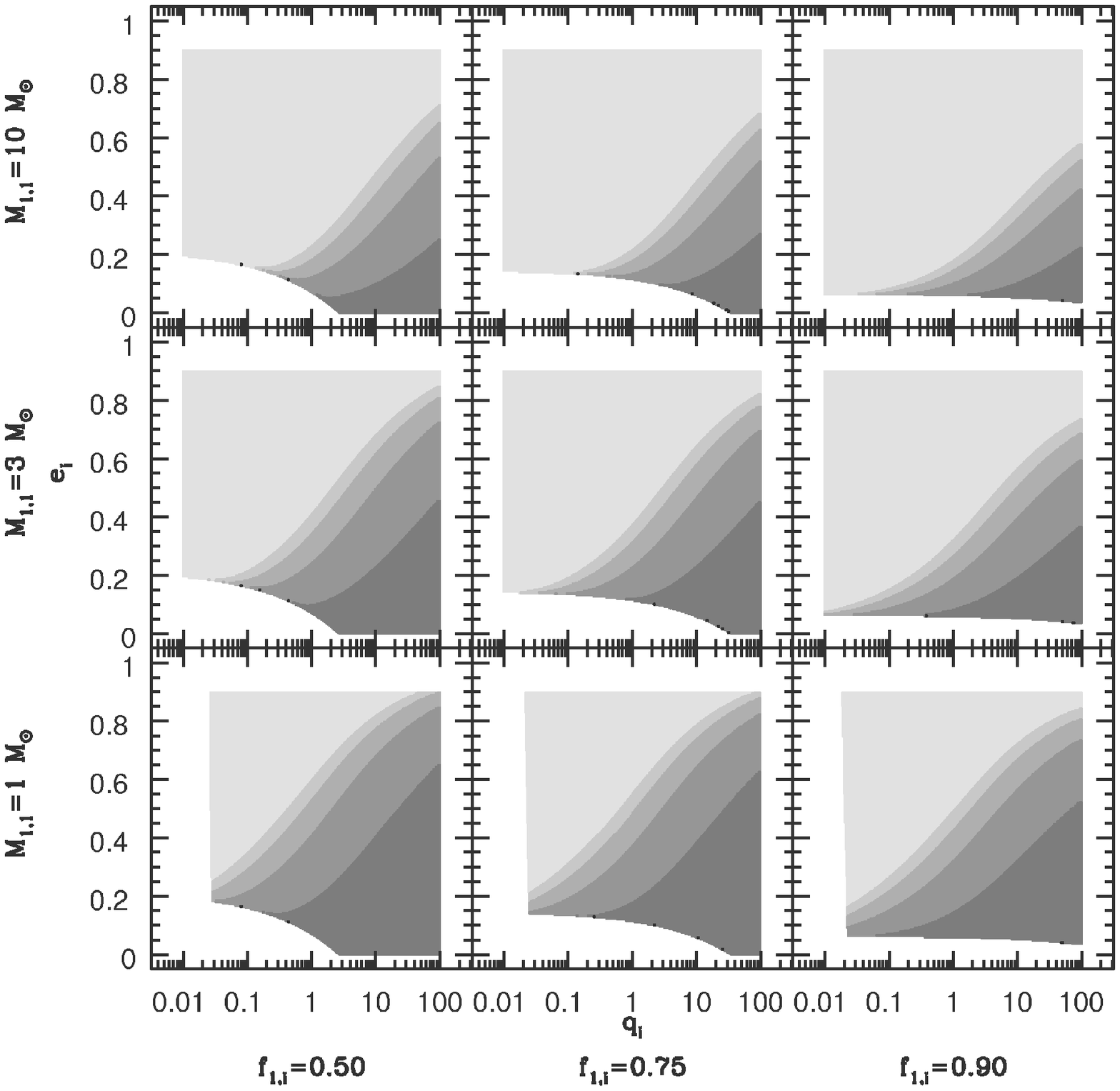}
\caption{Same as Figure~\ref{fig-dadtSA}, but for the evolution 
of the orbital eccentricity $e$.}
\label{fig-dedtSA}
\end{figure*}

In Figures~\ref{fig-dadtDI}\,--\,\ref{fig-dedtSA}, we show the 
variations of the instantaneous characteristic timescales $\tau_a =
|a/\dot{a}|$ for the evolution of the orbital semi-major axis $a$ 
and $\tau_e = |e/\dot{e}|$ for the evolution of the orbital eccentricity 
$e$, as contour plots in the $(q_i,e_i)$-plane at the start of mass 
overflow. The considered range of initial mass ratios is $0.01 \leq q_i 
\leq 100$. Each panel in the figures corresponds to different values of 
the initial donor mass $M_{1,i}$ and the initial rotation rate 
$f_{1,i}$. The mass overflow rate at periastron is taken to be 
$\dot{M}_0 = -10^{-9}\,M_\odot\, {\rm yr^{-1}}$. Since the resulting 
timescales are inversely proportional to $\dot{M}_0$, the results can 
easily be rescaled to different mass overflow rates. The radii of the 
two stars are calculated using the analytic formulae for stellar 
evolution derived by \citet{2000MNRAS.315..543H}.

To calculate the instantaneous rates of secular evolution of the orbital 
elements, we follow the evolution of the orbital semi-major axis and 
eccentricity of eccentric, effective Roche lobe overflowing binaries for 
100 
consecutive orbits. During this evolution, transitions may take place 
from direct impact accretion to self-accretion or vice versa. Systems 
for which this happens within 100 orbits from the start of mass overflow 
are confined to a fairly restricted region of the parameter space and 
are shown in black in Figures~\ref{fig-dadtDI}--\ref{fig-dedtSA}. In 
addition, for small initial mass ratios $q_i \la 0.01-0.02$, the radius 
of star~2 is so large relative to the size of the orbit that the binary 
becomes a contact binary rather than a semi-detached binary. These 
binaries are also left out of consideration in 
Figures~\ref{fig-dadtDI}--\ref{fig-dedtSA}. For binaries that do not 
transition between direct impact and self-accretion and that do not form 
contact binaries, timescales of orbital evolution are calculated by 
averaging the changes in the orbital elements over one orbital period.

Figures~\ref{fig-dadtDI} and~\ref{fig-dedtDI} show the orbital evolution 
timescales for the semi-major axis and eccentricity for the case of 
direct impact accretion. The dotted line in the figures separates 
regions of the parameter space where $a$ and $e$ increase (to the left 
of the dotted line) from regions of the parameter space where $a$ and 
$e$ decrease (to the right of the dotted line). For small orbital 
eccentricities, the transition from increasing to decreasing orbital 
semi-major axis occurs near $q_i=1$, consistent with what is expected 
for conservative mass transfer in binaries with circular orbits (see, 
e.g., Paper~I). As the eccentricity increases, the dividing line between 
increasing and decreasing semi-major axis moves toward smaller $q_i$. 
For a given choice of binary parameters, the orbital eccentricity 
typically evolves on shorter timescales than the orbital semi-major 
axis. The eccentricity evolution furthermore tends to circularize the 
orbit on fairly short timescales for most of the explored parameter 
space. Eccentricity pumping is only found for low initial binary mass 
ratios ($q_i \la 0.1$), low initial orbital eccentricities ($e_i \la 
0.05$), and substantially subsynchronous rotation ($f_{1,i} \la 0.75$).

Figures~\ref{fig-dadtSA} and~\ref{fig-dedtSA}, show the orbital 
evolution timescales for the orbital semi-major axis and eccentricity in 
the case of self-accretion. Unlike the direct impact case, 
self-accretion always acts to decrease both the semi-major axis $a$ and 
the orbital eccentricity $e$. The evolution of both $a$ and $e$ is 
furthermore fastest for larger binary mass ratios and lower orbital 
eccentricities. Comparison of Figures~\ref{fig-dadtSA} 
and~\ref{fig-dedtSA} with Figures~\ref{fig-dadtDI} and~\ref{fig-dedtDI} 
furthermore shows that the contours for self-accretion systems 
complement those for direct impact systems.

\subsection{Example Orbital Evolution Sequences}
\label{sec-sequences}

\begin{figure*}
 \plottwo{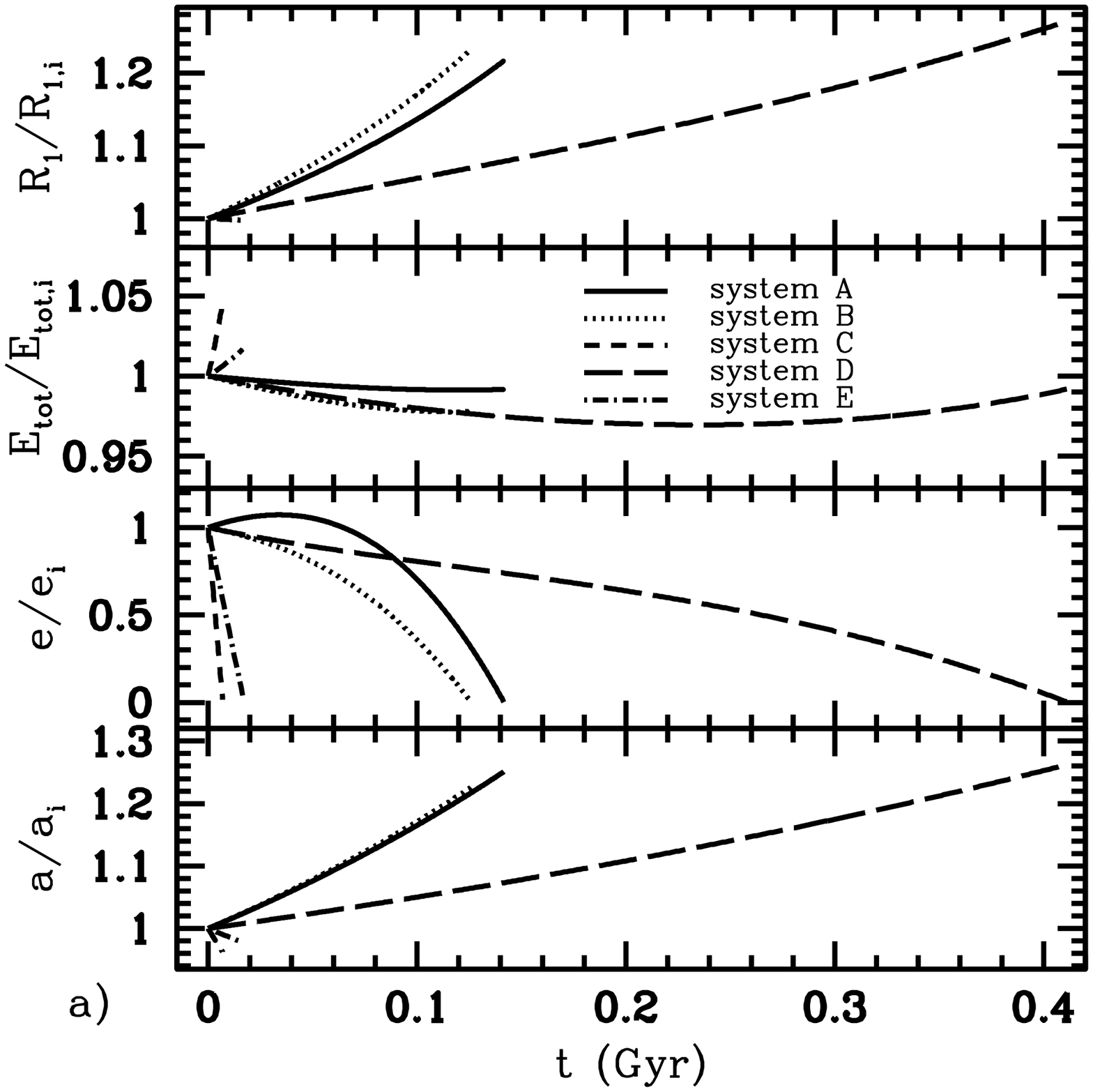}{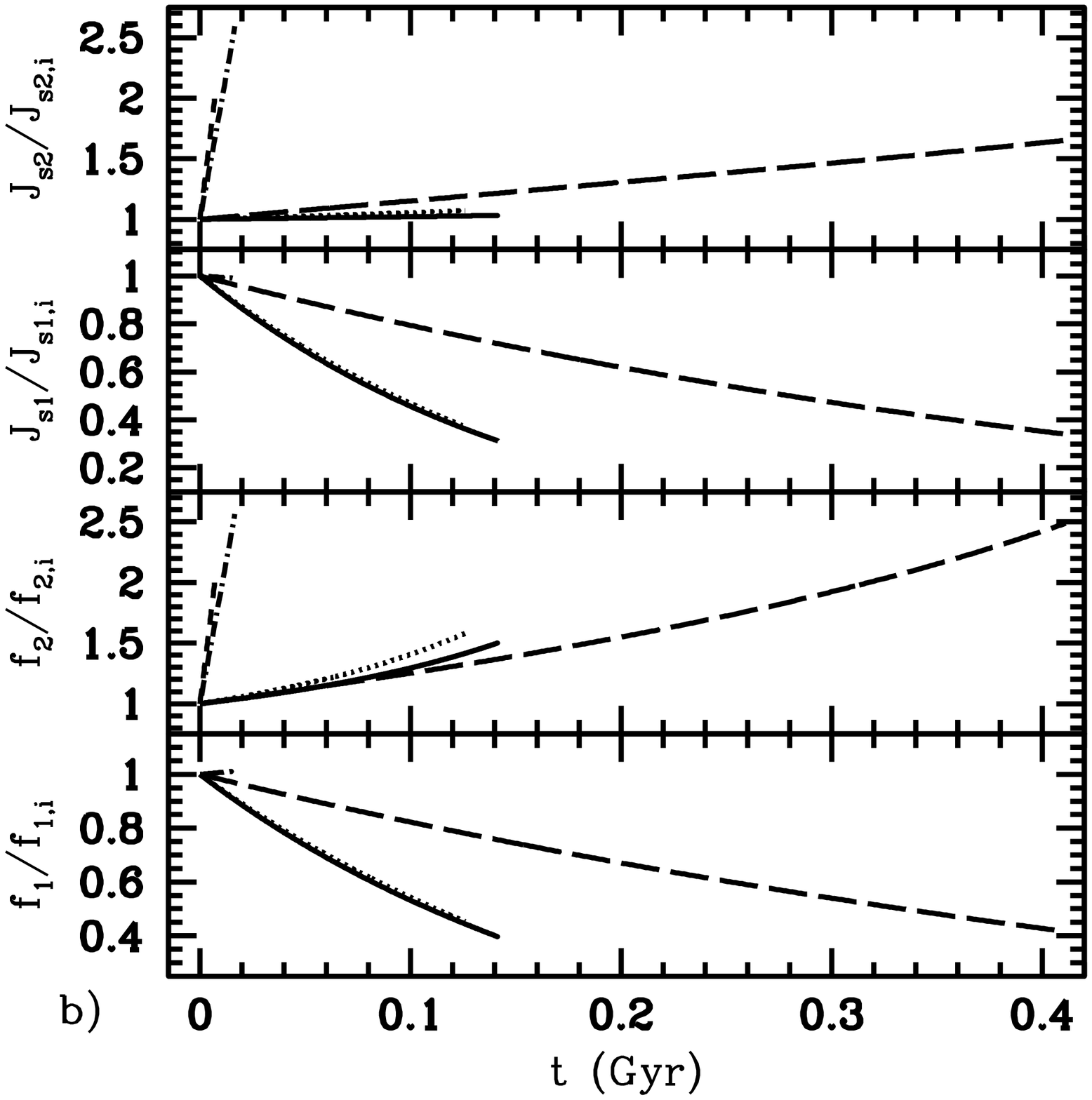}
 \caption{ Evolution of orbital parameters as a function of time for 
 systems undergoing direct impact accretion.  The $y$-axes show the 
 fractional change in the parameters relative to their initial value.  
 The initial and final values for some of the system parameters are 
 shown in table~\ref{tab-sys}. In {\bf (a)}, we show, from top to 
 bottom, the radius of star~1, the total energy of the system (orbital 
 plus the spin of each component), the eccentricity, and the semi-major 
 axis. In {\bf (b)}, we show, from top to bottom, the spin angular 
 momentum of star~2, the spin angular momentum of star~1, and the 
 rotation rates of stars~2 and 1.}
 \label{fig-diev}
\end{figure*}

\begin{figure*}
 \plottwo{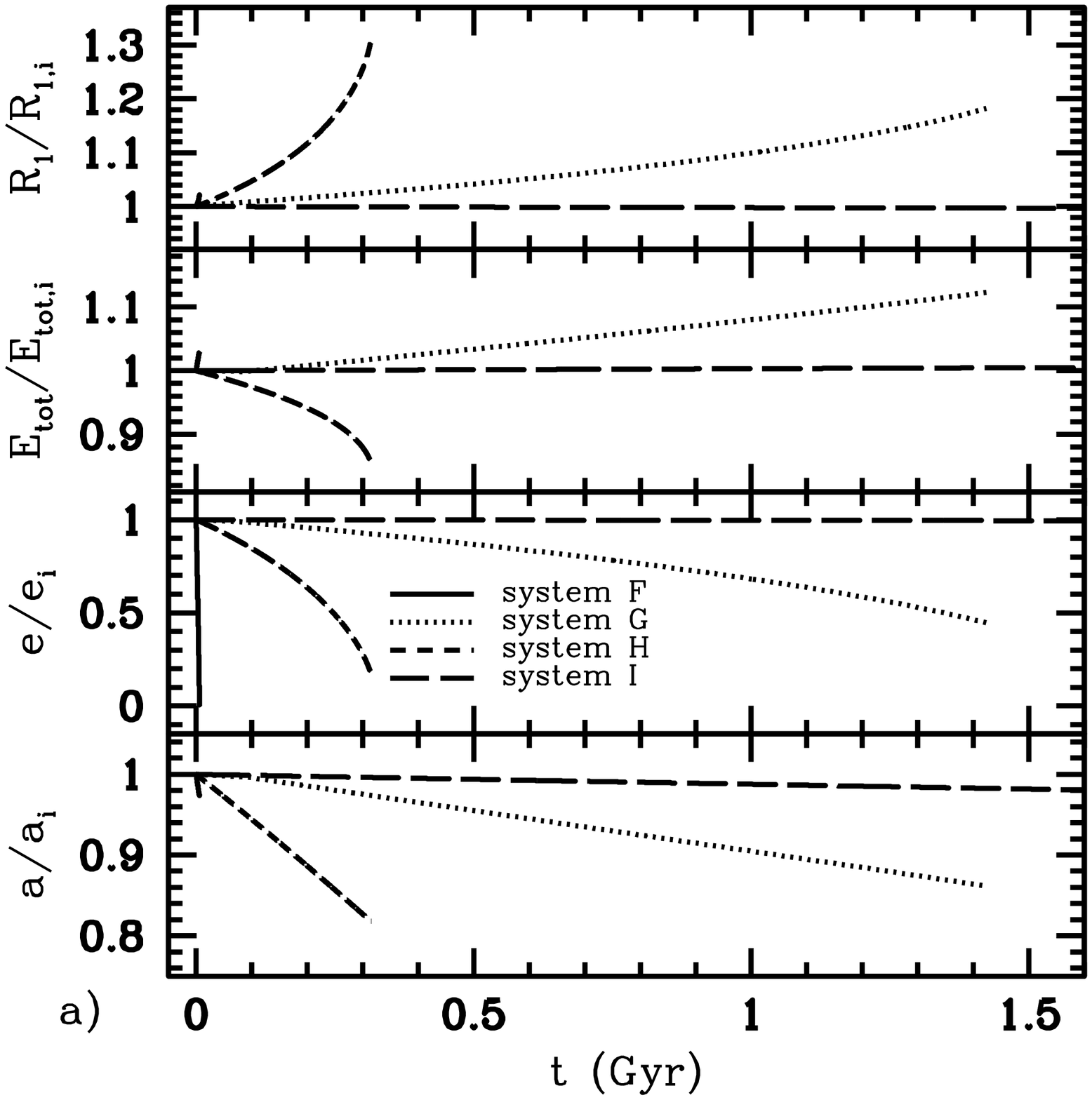}{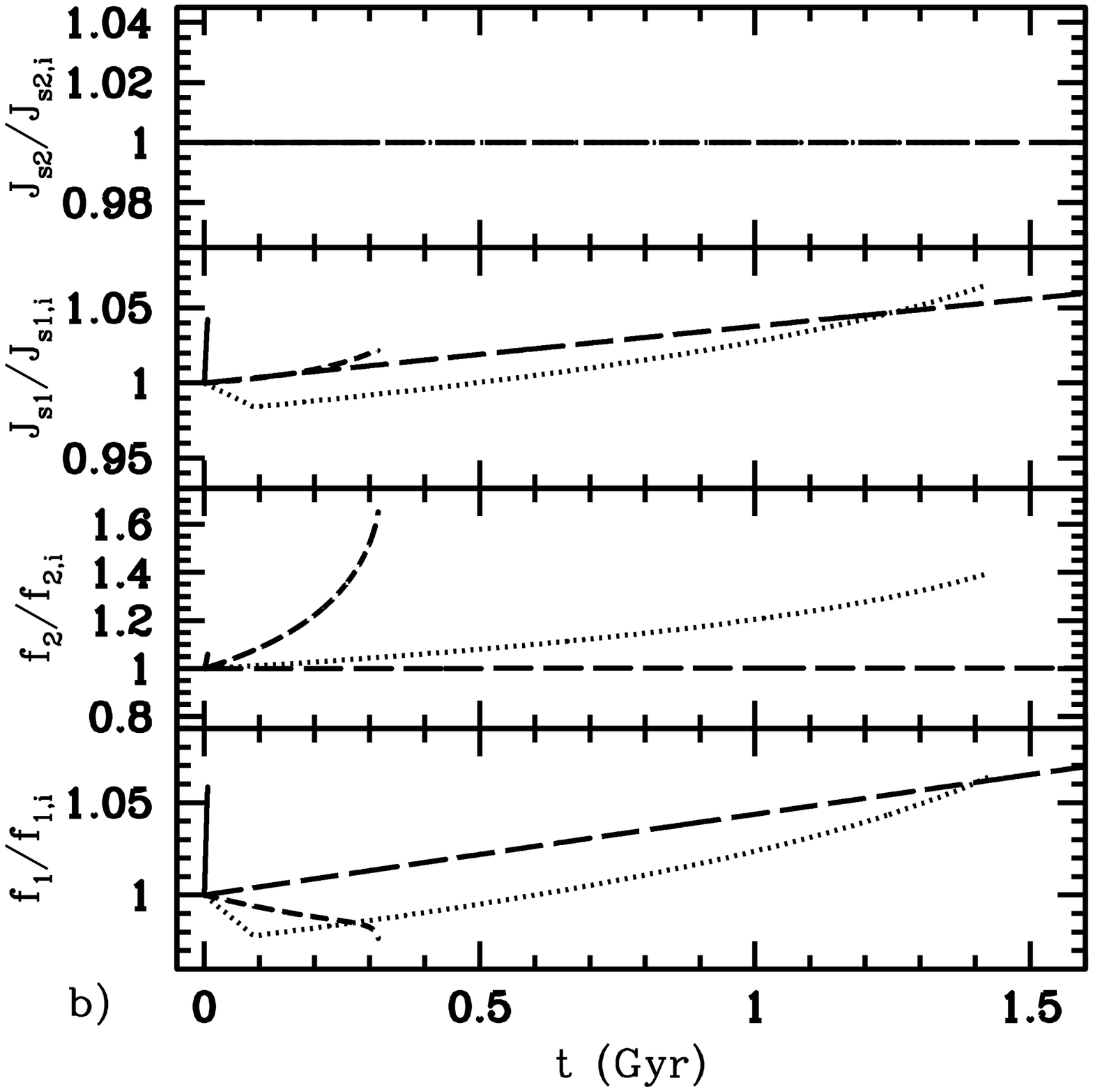}
 \caption{Same as Figure~\ref{fig-diev}, but for systems undergoing 
 self-accretion.}
 \label{fig-saev}
\end{figure*}

In Figures~\ref{fig-diev} and \ref{fig-saev} we show the long-term 
evolution of orbital and stellar properties for a representative sample 
of systems undergoing direct impact and self-accretion, respectively, 
for a mass overflow rate of $\dot{M}_0 = -10^{-9}\,M_\odot\,{\rm 
yr^{-1}}$.  Following the procedure outlined in \S\ref{sec-MTinecc}, a 
particle is ejected from the donor star at each periastron passage of 
the binary, and its ballistic orbit is followed until direct impact or 
self-accretion occurs. Impact of the particle on either star is 
determined under the assumption that the star is spherically symmetric
 and uniform in density ($I=0.4M{\cal R}^2$).
Hence, we do not account for deviations from spherical symmetry due to 
rotation and tidal interactions.

Following the accretion of the particle, the system evolves as a two 
body system until it reaches the periastron of the new orbit. At this 
point, the effective Roche lobe radius will also have changed compared 
to its value at the time of the previous mass ejection due to changes in 
the binary mass ratio, semi-major axis, eccentricity, and donor rotation 
rate. Since we assume the radius of the donor star to remain equal to 
the radius of its effective Roche lobe at periastron during the lifetime 
of the mass overflow phase, the donor rotation rate must be updated in 
accordance with conservation of spin angular momentum before the next 
mass ejection at periastron. Because the donor rotation rate in turn 
affects the radius of the effective Roche lobe, a root-finding algorithm 
is used to calculate a new self-consistent radius and rotation rate for 
the donor star immediately before the ejection of matter at each 
consecutive periastron passage.  Throughout our numerical integrations, 
the total angular momentum of the system (defined as the spin angular 
momenta of stars~1 and 2 plus the orbital angular momentum of the 
system), is conserved to within a factor of $10^{-7}$.

\begin{deluxetable*}{cccccccccccccccc}
\tablecolumns{16}
\tablewidth{0pt}
\tablecaption{Initial and final binary parameters for the evolutionary 
sequences shown in Figures~\ref{fig-diev} and~\ref{fig-saev}}
\tablehead{
\colhead{} & \multicolumn{2}{c}{$M_1\,(M_\odot)$} & \multicolumn{2}{c}{$M_2\,(M_\odot)$} & \multicolumn{2}{c}{$q$} &\multicolumn{2}{c}{$a\,(R_\odot)$} & 
\multicolumn{2}{c}{$e$} & \multicolumn{2}{c}{$f_1$} & 
\multicolumn{2}{c}{$f_2$} &\colhead{$t_{\rm final}$}\\
\\
\colhead{System} & \colhead{Initial} & \colhead{Final} & \colhead{Initial} & \colhead{Final} & \colhead{Initial} & \colhead{Final} &\colhead{Initial} & \colhead{Final} & 
\colhead{Initial} & 
\colhead{Final} & \colhead{Initial} & \colhead{Final} & \colhead{Initial} & \colhead{Final} &\colhead{(Gyr)}}
\startdata
A & 1.00  & 0.78 & 20.00 & 20.02  & 0.05 & 0.04 & 9.31  & 11.65 & 0.025 & 0.000 & 0.50 & 0.20 & 1.00 & 1.50 & 0.14\\
B & 1.00  & 0.81 & 10.00 & 10.19  & 0.10 & 0.08 & 7.85  & 9.63  & 0.050 & 0.000 & 0.50 & 0.23 & 1.00 & 1.58 & 0.13\\
C & 1.00  & 0.99 &  0.25 &  0.26  & 4.00 & 3.81 & 3.24  & 3.12  & 0.050 & 0.000 & 0.75 & 0.75 & 1.00 & 2.00 & 0.01\\
D & 3.00  & 2.38 & 15.00 & 15.62  & 0.20 & 0.15 & 2.26  & 28.56 & 0.050 & 0.000 & 0.75 & 0.31 & 1.00 & 2.49 & 0.41\\
E & 10.00 & 9.97 &  1.00 &  1.03  & 10.00& 9.68 & 19.78 & 19.32 & 0.025 & 0.000 & 0.90 & 0.91 & 1.00 & 2.57 & 0.02\\
\cline{1-16} \\
F & 1.00  &  1.00 & 0.25  & 0.25  & 4.00 & 4.00 & 3.16  & 3.07  & 0.050 & 0.000 & 0.50 & 0.53 & 1.00 & 1.06 & 0.01\\
G & 3.00  &  3.00 & 5.00  & 5.00  & 0.60 & 0.60 & 2.71  & 2.41  & 0.400 & 0.234 & 0.75 & 0.78 & 1.00 & 1.39 & 1.42\\
H & 3.00  &  3.00 & 0.50  & 0.50  & 6.00 & 6.00 & 17.41 & 14.23 & 0.400 & 0.063 & 0.90 & 0.88 & 1.00 & 1.66 & 0.32\\
I & 10.00 & 10.00 & 100.0 & 100.0 & 0.10 & 0.10 & 266.55& 253.65& 0.800 & 0.789 & 0.50 & 0.58 & 1.00 & 1.02 & 11.9 
\enddata
\tablecomments{$M_1$ is the mass of the donor star, $M_2$ the mass of 
the companion, $q=M_1/M_2$ the mass ratio, $a$ the semi-major axis, $e$ 
the orbital eccentricity, $f_1$ the rotational angular velocity of the 
donor star in units of the orbital angular velocity at periastron, $f_2$ 
the rotational angular velocity of the donor star in units of the 
orbital angular velocity at periastron,  and $t_{\rm final}$ the time at 
which the calculation was terminated.}
\label{tab-sys}
\end{deluxetable*}

In Figures~\ref{fig-diev} and~\ref{fig-saev}, the fractional changes of 
the donor radius ${\cal R}_1$, total energy, $E_{\rm tot}$, 
eccentricity, $e$, semi-major axis, $a$, spin orbital angular momentum 
of stars~1 and 2, $J_{s1}$ and $J_{s2}$, and the rotation rates of 
stars~1 and 2, $f_1$ and $f_2$, are shown as a function of time for 
systems undergoing direct impact accretion and self-accretion, 
respectively.  The evolutionary sequences were calculated for up to four 
mass transfer timescales $M_1/|\dot{M}_0|$ (systems G and I) or until 
the binary circularized (systems A through F), whichever occurred first. 
For system H, the calculation was terminated at time $t = 0.32\,{\rm 
Gyr}$ when it ceased to be a self-accretion system. The initial and 
final parameters of the nine evolutionary sequences shown are summarized 
in Table~\ref{tab-sys}.

The direct impact systems shown in Figure~\ref{fig-diev} all circularize 
within a few Gyr or less after the start of mass overflow. In the case 
of system~$A$, the eccentricity initially grows, reaching a peak $e_{\rm 
max}=0.0268$ at $t = 0.034$\,Gyr, before circularizing at an 
increasingly rapid rate.  The semi-major axes generally increase, except 
for systems C and E, as expected for binaries with mass ratios $q>1$ and 
almost circular orbits. Mass loss from the donor star and the growth of 
the donor's radius, which is required to track the growing effective 
Roche lobe, furthermore decrease the donor's rotation rate to 
significantly sub-synchronous values at the end of the evolutionary 
sequence for systems A, B, and D. For systems C and E, mass transfer to 
the less massive star~2 causes the semi-major axis to decrease, 
increasing the orbital angular velocity at periastron, and thereby 
effectively {\it increasing} the donor rotation rate.  This in turn 
causes the effective Roche lobe radius of the donor to shrink. We note 
that in all five cases shown, the end of the evolutionary sequence 
corresponds to the time at which the binary becomes circular, not to the 
cessation of mass transfer in the binary system.

The self-accretion systems shown in Figure~\ref{fig-saev} decrease their 
eccentricity as a result of mass overflow, though system I has an 
evolutionary timescale for the orbital eccentricity much longer than a 
Hubble time (see Figure~\ref{fig-dedtSA}). The orbital semi-major axes 
always decrease, while the donor rotation rates can increase as well as 
decrease during the course of mass overflow. There is no change in the 
spin angular momentum of star~2 since there is no mass transferred.  In 
this case, the change in $f_2$ is due entirely to the change in the 
orbital angular velocity at periastron. In the case of system~$G$, the 
donor initially spins down by a few per cent relative to the rotational 
angular velocity at periastron, but reverses its spin evolution after 
about 100\,Myr. At the end of the evolutionary phase shown 
(corresponding to four mass transfer time scales), the donor is spinning 
super-synchronously at the periastron of the binary orbit.  The donor 
rotation rate for System~$H$ slowly decreases until just past $t \approx 
0.3$\,Gyr, at which time it rapidly plunges to very low rotation rates. 
We note that the physical rotation rate of this donor star never 
actually decreases, and that the decrease in $f_1$ is caused by the 
orbital angular velocity at periastron increasing at a rate faster than 
the actual rotation rate of the donor star. The rapid decrease starts a 
few orbits before the system transitions from a self-accretion system to 
a system in which the ejected matter impacts neither star~1 nor star~2 
during the course of one binary orbit. Since our code is currently not 
equipped to deal with the evolution of systems like this, we terminated 
the calculation at this point.

We note that at every step of the orbital evolution 
calculations, both the total linear and angular momenta of the system 
remain constant. However, mass overflow can exchange momentum between 
the spins of the binary components and the orbit. We particularly find 
that, contrary to past assumptions \citep[for example, 
see][]{1988ApJ...332..193V, 2004MNRAS.350..113M}, direct impact 
accretion does not necessarily provide a sink of orbital angular 
momentum \citep[see also][]{2007ApJ...670.1314M}. Instead, the 
transferred matter contains both spin and orbital angular momentum from 
the donor star, part of which can be returned to the orbit upon 
accretion. This makes it possible for the orbital angular momentum to 
increase at the expense of the spin angular momentum of the donor. The 
impact of this result on the stability of mass overflow in close 
binaries such as double white dwarfs will be the subject of a 
forthcoming investigation.

Lastly, we can see from Figures~\ref{fig-diev} and~\ref{fig-saev} that 
the total energy of the system can change significantly over the course 
of its lifetime.  Due to our assumption that the particle is ejected and 
accreted perfectly inelastically (i.e., with no loss of momentum), it is 
impossible to conserve the total system energy as well.  Even in the 
case of self-accretion where the mass of star~1 is unchanged the ejected 
particle self-accretes with a different position and velocity than that 
with which is was ejected.  For the systems tested here, the total 
energy can change by as much as $10\%$ or more, and can either increase 
or decrease depending on the system parameters.  The energy added or 
subtracted from the system might be accompanied by a 
commensurate change in the thermal energy of the stars' envelopes.

\section{Discussion and Conclusions}

In this paper, we extended our previous work on mass transfer in 
eccentric binaries by using ballistic trajectory calculations to 
determine the parameter space for different outcomes of mass overflow. 
Assuming mass overflow takes place through Roche lobe overflow at the 
periastron of the binary orbit, we explored a broad parameter space to 
determine the conditions under which mass overflow leads to direct 
impact accretion onto the donor's companion or to fallback onto the 
donor star (``self-accretion''). The results presented in this paper 
along with those presented in \citet{2007ApJ...660.1624S, 
2007ApJ...667.1170S, 2009ApJ...702.1387S}, provide a self-consistent 
picture of the orbital evolution of eccentric effective Roche lobe 
overflowing binary star systems as a function of their initial orbital 
parameters. Until now, such a picture has been lacking both in binary 
stellar evolution and binary population synthesis codes.

We find that systems with a large initial eccentricity ($e_i \gtrsim 
0.2$) or a donor initially rotating near synchronicity ($f_{1,i} \gtrsim 
0.8$) are always expected to undergo self-accretion. Direct impact 
accretion or disk formation is expected to occur mainly for systems with 
low eccentricities, low mass ratios, and substantially subsynchronously 
rotating donor stars.  Furthermore, self-accretion is found to always 
decrease the orbital eccentricity, and can do so over timescales ranging 
from less than a Myr to more than a Gyr depending on the initial binary 
parameters and the chosen mass overflow rate.

To illustrate the applicability of the presented formalism, we 
calculated the orbital evolution due to mass overflow for eccentric 
binaries consisting of an evolved main sequence donor and a zero-age 
main sequence accretor as a function of the initial binary parameters. 
For binaries undergoing direct impact accretion, mass transfer can 
increase as well decrease the initial orbital semi-major axis and 
eccentricity, while for binaries undergoing self-accretion, mass 
overflow always decreases both the initial orbital semi-major axis and 
eccentricity. For a mass overflow rate of $\dot{M}_0=-10^{-9}\,M_\odot\, 
{\rm yr}^{-1}$, the time scales of orbital evolution can range from less 
than 1\,Gyr to more than a Hubble time.

In this exploratory study, both the donor star and the accretor were 
treated as rigid spheres. The results presented therefore do not account 
for orbital evolution due to tidal torques or for the radial response of 
the donor star to mass loss. Our calculations nevertheless clearly 
illustrate that mass overflow leading to direct impact accretion or 
self-accretion can both enhance or counteract the effects of tides in 
eccentric effective Roche lobe overflowing binaries. A similar 
conclusion was reached by \citet{2007ApJ...667.1170S, 
2009ApJ...702.1387S} for binaries in which mass transfer leads to the 
formation of an accretion disk. As one of the next steps in our 
investigation, we will implement more realistic stellar models in our 
ballistic particle trajectory code and examine the effects of tides and 
stellar evolution on mass transfer in eccentric binaries.

While not discussed in detail here, we also find that direct impact 
accretion does not necessarily deposit all of the accreted matter's 
angular momentum into the spin of the accretor. Instead, some of this 
angular momentum, which originated from both spin and orbital angular 
momentum of the donor star, is deposited into the accretor's spin while 
the remaining part is returned to the orbit. This is in stark contrast 
to commonly adopted assumptions in stability studies of mass transfer in 
ultra-compact binaries \citep[e.g.,][]{1988ApJ...332..193V, 
2004MNRAS.350..113M}. The possibility of avoiding strong orbital angular 
momentum loss during direct impact accretion is particularly relevant in 
predicting expected gravitational wave detection rates from 
mass-transferring double white dwarfs (AM\,CVn stars) by the Laser 
Interferometer Space Antenna (LISA).

\acknowledgements

We thank Christopher Deloye for numerous useful discussions, Francesca 
Valsecchi for the use of her C implementation of Steffen's 
(\citeyear{1990A&A...239..443S}) interpolation algorithm used to 
calculate the time derivatives of the orbital elements, and Christopher 
Tout for a number of useful suggestions.  This work is partially 
supported by a NASA Graduate Fellowship (NNG04GP04H/S1) to J.S., and a 
NSF CAREER Award (AST-0449558), a Packard Fellowship in Science and 
Engineering, and a NASA ATP Award (NAG5-13236) to V.K.

\bibliography{MT}

\end{document}